\documentclass[onecolumn,amsmath,amssymb,nofootinbib,11pt]{article}

\usepackage{jheppub}
\usepackage{ifpdf}

\usepackage[latin1]{inputenc}
\usepackage{graphicx,subcaption}
\usepackage{amsfonts}
\usepackage{amsmath}
\usepackage{amssymb}

\usepackage{epsfig}
\usepackage{pdfpages}
\usepackage{graphicx,epstopdf}
\usepackage[makeroom]{cancel}
\usepackage{hyperref}
\hypersetup{pdftex,colorlinks=true,allcolors=blue}
\usepackage{array}
\usepackage{tikz}
\usepackage{mathtools}
\usepackage{braket}
\usepackage{yhmath}

\newcommand{\RN}[1]{%
  \textup{\uppercase\expandafter{\romannumeral#1}}%
}

\newcommand{\del}{\partial}

\newcommand{\mt}[1]{\textrm{\tiny #1}}

\newcommand{\mC}{\mathcal{C}}

\newcommand{\Gn}{G_\mt{N}}

\newcommand{\sh}{\sinh}
\newcommand{\ch}{\cosh}
\newcommand{\no}{\nonumber}

\newcommand{\LB}{L_{\mathcal{B}}}

\newcommand{\bea}{\begin{eqnarray}}
\newcommand{\eea}{\end{eqnarray}}

\usepackage[thinlines]{easytable}

\graphicspath{{./figures/}}

\begin{document}

\preprint{LPTENS/18/16}
\title{Holographic Complexity for Defects Distinguishes Action from Volume}

\author[a]{Shira Chapman,}
\author[b]{Dongsheng Ge,}
\author[b]{Giuseppe Policastro,}
\affiliation[a]{Institute for Theoretical Physics, University of Amsterdam
Science Park 904, Postbus 94485, 1090 GL Amsterdam, The Netherlands}
\affiliation[b]{Laboratoire de Physique Th\'eorique de l'\'Ecole Normale Sup\'erieure, CNRS, 
Universit\'e PSL, Sorbonne Universit\'es, Universit\'e Pierre et Marie Curie, 24 rue Lhomond, 75005 Paris, France}

\emailAdd{s.chapman@uva.nl}
\emailAdd{ge@lpt.ens.fr}
\emailAdd{policast@lpt.ens.fr}

\date{\today}

\date{\today}

\abstract{We explore the two holographic complexity proposals for the case of a  2d boundary CFT with a conformal defect. We focus on a Randall-Sundrum type model of a thin AdS$_2$ brane embedded in AdS$_3$. We find that,  using the  ``complexity=volume'' proposal, the presence of the defect generates a logarithmic divergence in the complexity of the full boundary state with a coefficient which is related to the central charge and to the boundary entropy. For the ``complexity=action'' proposal we find that the complexity is not influenced by the presence of the defect. This is the first case in which the results of the two holographic proposals differ so dramatically. We consider also the complexity of the reduced density matrix for subregions enclosing the defect. We explore two bosonic field theory models which include two defects on opposite sides of a periodic domain. 
We point out that for a compact boson, current free field theory definitions of the complexity would have to be generalized to account for the effect of zero-modes.}

	\maketitle

\section{Introduction}

In the last few years a lot of effort has been devoted to understanding the relation between certain properties of complex quantum systems (such as thermalization, scrambling, and chaos) and their counterparts on the holographically dual gravitational side. The quest for such relations dates back to the discovery of the holographic duality in 1997 \cite{Maldacena:1997re}, but a more recent facet of this story is the incorporation of quantum information-theoretic aspects on top of more traditional physics considerations.

Thinking of black holes as quantum computers, along the lines first advocated by \cite{Hayden:2007cs}, Susskind and collaborators have argued that some of the puzzles associated to the region behind the horizon of black holes could be clarified if one considers a quantity associated to the quantum state, that they named complexity, borrowing a notion from quantum information, see, e.g., \cite{Susskind:2014rva,Susskind:2014moa}. The complexity of a quantum state (say, of a finite number of qubits) is defined, with respect to a given reference state, as the minimum number of gates (i.e., unitary operators, taken from some universal set) that need to be applied to a certain reference state $\ket{R}$ to approximate a desired target state $\ket{T}$ \cite{Aaronson:2016vto,Encyclopedia}.
The main point of Susskind's argument was that the complexity captures properties of quantum states which are not encoded in previously studied quantities such as the entanglement or R\'enyi entropies. In particular if we consider a simple state that has not yet reached its maximal complexity and evolve it with a generic fast scrambling Hamiltonian then the complexity will always tend to increase \cite{Brown:2017jil}, just as entropy does, but whereas entropy saturates its growth after thermalization \cite{Hartman:2013qma}, complexity is expected to keep growing for an exponentially longer time. This property is reflected holographically in the growth of the area of a Cauchy surface that connects the two boundaries of the two-sided eternal black hole, dual to the thermofield double state \cite{Maldacena:2001kr}, going across the horizon.

These considerations led to the proposal that the volume of the Cauchy slice in the eternal black hole would be the precise holographic counterpart of the complexity of the thermofield double state in the dual theory. More precisely, the Complexity=Volume (CV) conjecture \cite{Susskind:2014rva,Stanford:2014jda} posits that
\begin{equation}\label{eq:CVformula}
\mC_V = \frac{V}{\Gn L},
\end{equation}
where $V$ is the volume of a maximal codimension-one hypersurface that ends on the boundary time slice where the state is defined, and $\Gn$ is the Newton constant. One of the drawbacks in the CV conjecture is that a length scale, here $L$, has to be introduced manually for dimensional reasons. In order to cure this problem, a second conjecture, the  Complexity=Action (CA) was proposed  by Brown, Susskind et al. \cite{Brown:2015bva,Brown:2015lvg}, in which
\begin{equation}\label{eq:CAformula}
\mC_A ={  I_{\mt{WDW}} \over \pi\hbar},
\end{equation}
where $I_{\mt{WDW}}$ is the on-shell gravitational action evaluated on the Wheeler-DeWitt (WDW) patch\footnote{The WDW patch is the bulk causal development of the slice from the CV conjecture, and is bounded by light sheets sent from the boundary time slice where the state is defined. For the case of the two-sided AdS black hole the WDW patch is anchored at both boundaries at the relevant times where the thermofield double state is studied.} and we will set $\hbar=1$ from now on. The latter proposal is seemingly more universal, since the particular maximal slice does not play a preferred role.
Technically speaking, the action approach is more subtle than the volume one, since codimension-1 boundary surface terms and codimension-2 joint terms have to be included in the action, see \cite{Gibbons:1976ue,York:1972sj,Parattu:2015gga,Hayward1993,Brill:1994mb,Lehner:2016vdi} (a summarized prescription is outlined in appendix A of \cite{Carmi:2016wjl}).

Various aspect of the two holographic proposals have been explored in recent years. In particular, their structure of divergences \cite{Comments,Reynolds:2016rvl}, their time dependence \cite{Susskind:2014rva,Brown:2015bva,Brown:2015lvg,Lehner:2016vdi,Carmi:2017jqz} and their reaction to shockwaves \cite{Stanford:2014jda,Chapman:2018dem,Chapman:2018lsv,Zhao:2017iul}. In general, it appears that the predictions of the action and volume proposals tend to coincide up to overall numerical factors. For instance, the complexity grows linearly for a long period of time at a rate which is proportional to the energy of the system.\footnote{For the volume this statement holds in the high temperature limit, and for hyperscaling violating geometries \cite{Alishahiha:2018tep,Swingle:2017zcd} the late time growth rate from the CV proposal includes an additional  temperature dependent proportionality coefficient required from dimensional analysis considerations.} In shockwave geometries the complexity exhibits characteristic delays in its growth related to the scrambling time of the system both for the CV and CA proposals. It is therefore of interest to extend the study of complexity to systems where a clear-cut distinction can be made between the predictions of the action and the volume proposals.

In order to explore this question, we consider the modification of the complexity associated to the introduction of a conformal defect in the field theory. Defect CFTs (DCFTs) have been studied extensively in the literature (see, e.g., \cite{Andrei:2018die} and references therein), both on the field theory side and holographically, so we can draw on existing constructions. For simplicity, we focus on the case of a 2d DCFT, and consider its ground state complexity.
On the gravitational side, we consider a bottom-up Randall-Sundrum type model \cite{Aharony:2003qf} of a thin AdS$_2$ brane embedded in AdS$_3$ spacetime \cite{Azeyanagi:2007qj}. The brane, which acts as a defect in this geometry, has two anchoring points on the boundary, which introduce defects in the boundary theory on opposite sides of the circular domain  (analogous to a quark-antiquark pair).
The brane backreacts and modifies the geometry, therefore entailing a modification of the complexity depending on a parameter, namely the tension of the brane.

Our main results can be found in eqs.~\eqref{eq:CVresult} and \eqref{final-result-CA}, for the CV and CA proposals,  respectively. In particular we observe that the result of the CA proposal does not depend on the tension of the  brane, so the complexity appears unaffected by the introduction of the defect while for the CV proposal a new logarithmic divergence appears in the complexity due to the presence of the defect which is related to the central charge and to the Affleck-Ludwig boundary entropy \cite{Affleck:1991tk}.\footnote{The system with the defect is related to a similar system with a boundary via a folding trick.} It is worth noting that this is the first case  in which the results of the holographic CV and CA proposals disagree so dramatically. This offers an opportunity to discriminate between the two prescriptions, which appear to be computing different quantities.\footnote{Previous studies of the CV complexity in the presence of boundaries can be found in \cite{Flory:2017ftd} where a holographic Kondo model was explored and it was pointed out that for constant tension branes the complexity increases with the tension of the brane. We reach a similar conclusion in our model using the CV proposal. For an alternative proposal of world sheet complexity, see section 4 of \cite{deBoer:2017xdk}.}

In the original proposals, the complexity is associated to the state of the whole system.
Inspired by the Ryu-Takayanagi prescription for the holographic entanglement entropy \cite{Ryu:2006bv,Ryu:2006ef}, and motivated by the suggestion that the reduced density matrix of a boundary subregion is encoded in its ``entanglement wedge'' \cite{Czech:2012bh,Headrick:2014cta},\footnote{The entanglement wedge is defined as the set of bulk points that are spacelike separated from the RT surface and causally connected to the causal wedge on the boundary of AdS.}
proposals have been made \cite{Alishahiha:2015rta,Carmi:2016wjl} for an extension of the complexity conjectures for states (reduced density matrices) associated to subregions. For static geometries, the CV prescription is generalized to the volume enclosed between the RT surface and the AdS boundary, while for the CA prescription, one considers the gravitational action of the region enclosed between the WDW patch and the entanglement wedge.
We consider the subregion complexity for the defect geometry for subregions which include a single defect; For the CA proposal we are able to perform the computation only in the case of a symmetric region, i.e., when the defect is at its midpoint. The results are in eq.~\eqref{eq:subCVfinal} for the CV proposal and in eqs.~\eqref{eq:subCAdresult}, \eqref{toderive} for CA proposal. Again we find no contribution in CA complexity from the defect. Interestingly, we find that the structure of divergences of the subregion CA complexity is not the same as the one of the total CA complexity. In particular, we observe a $\ln \Lambda$ divergence where $\Lambda$ is the UV momentum cutoff.

The evidence for the validity of the CV and CA proposals is far from being conclusive since the notion of complexity in QFT is still not well understood. Some progress has been made
\cite{Jefferson:2017sdb,Chapman:2017rqy,Yang:2017nfn,Kim:2017qrq,Hashimoto:2017fga,Khan:2018rzm,Hackl:2018ptj,
Reynolds:2017jfs,Alves:2018qfv,Camargo:2018eof,Caputa:2017urj,Caputa:2017yrh,Caputa:2018kdj,
Bhattacharyya:2018wym,Magan:2018nmu,Bhattacharyya:2018bbv,Takayanagi:2018pml,Ali:2018fcz,Chapman:2018hou}, but a precise  definition in strongly interacting CFTs, from first principles, is still absent. In particular, one can put together an operative definition for the case of free fields \cite{Jefferson:2017sdb,Chapman:2017rqy}, and arrive at a result that matches with holography in terms of the divergence structure. However, it is not clear at the moment what is the universal content which can be extracted from the coefficients of the complexity as a series expansion in the cutoff scale, similarly to the case of entanglement entropy, where the coefficient of the logarithmic divergence is associated to the central charge in the CFT. An operative definition for subregion complexity is also absent, although some proposals have been made in \cite{Agon:2018zso}.

The definition of complexity in QFT is subject to many ambiguities. In particular, one is free to choose a reference state as well as a set of gates that can act on the state. It was initially suggested \cite{Jefferson:2017sdb,Chapman:2017rqy}  that the ambiguity associated with the reference state is mirrored by a similar ambiguity on the gravitational side of the correspondence, related to the choice of  normalization of the null normals at the boundaries of the WDW patch. More recently it has been understood that in evaluating the complexity one has to include a certain counter term needed to restore reparametrization invariance on the null boundaries \cite{Chapman:2018dem,Chapman:2018lsv}. This counter term comes accompanied with a length scale which could be the one reproducing the effect of the various ambiguities on the complexity. For a more elaborate discussion, see section 5 of \cite{Chapman:2018lsv}.

We make a naive attempt to match our holographic results with the dual field theory.
As mentioned before, the definition of complexity for a generic field theory is unknown, and furthermore, the precise CFT dual of our holographic setup is not known. However, we can look at free field models of CFTs with defects, analogous to our holographic setup and study their complexities. We consider first  a model with permeable domain walls \cite{Bachas:2001vj} on opposite sides of a periodic domain. We observe in this case that a logarithmic contribution proportional to the parameters of the defect is absent, similarly to what happens for the CA proposal.
We then briefly discuss a solvable model with a boundary interaction \cite{Callan:1994ub}. It seems that in this case a logarithmic contribution which depends on the strength of the boundary interaction is present in the result for the complexity (in the case of a single boundary, but not if there are two boundaries), though we have to make an assumption that the formula derived for free fields, which computes the complexity in terms of the (single particle) spectrum, can be extended to these cases. An extension of the current holographic calculation to a case where the dual field theory is known would be required in order to identify which one of the field theory models (if any) is relevant for the analogy with holography. Since we are focusing on models of compact bosons, we point out that the effects of zero modes could have an influence on the complexity and might need to be incorporated into the existing definitions.

The paper is organized as follows: In Section \ref{sec:prelim}, we describe the defect AdS$_3$ geometry, employing different coordinate systems. We also discuss the choice of cutoff and the shape of the WDW patch in this geometry. In Section \ref{sec:CVCA},  we  present the calculations of the holographic complexity of the full boundary state using  both the CV and CA proposals.  In section \ref{sec:subregion},  we consider the subregion complexity proposals for subregions including one defect. In section \ref{sec:QFT}, we describe a free bosonic field theory model with defects as well as an exactly solvable model with a boundary interaction and compute their complexities. We conclude with a summary of the main results and a discussion in Section \ref{Discussion}. A number of technical details of the calculation are discussed in appendixes \ref{app:nullGeoApp} and \ref{app:sCAallpieces}, and the subregion CV proposal  in the Poincar\'e patch with two distinct cosmological constants on the two sides of the defect is discussed in appendix \ref{app:poincare}.

\section{Preliminaries}\label{sec:prelim}
In the following section we provide various ingredients of the defect toy model which we use to study the complexity of defects in this paper. As mentioned in the introduction, we focus our attention on a Randall-Sundrum solution for a 2d brane of tension $\lambda$ embedded in a 3d geometry which is a solution of Einstein equations with negative cosmological constant. Various aspects of this simple solution were already studied in \cite{Azeyanagi:2007qj}. We start by reviewing the solution in a number of convenient coordinate choices. We then address the choice of cutoff surface and describe how to  obtain null geodesics emanating from a point on the boundary in order to construct the WDW patch for the CA proposal.

\subsection{Two-Dimensional Branes in AdS$_3$}

We begin by considering the solution for a symmetric defect in AdS$_3$ which solves Einstein equations for the action
\begin{equation}\label{eq:gravaction}
S = \frac{1}{16 \pi \Gn} \int d^3 x \sqrt{-g} \left(R+\frac{2}{L^2} \right)-\lambda \int d^2 x \sqrt {-h} ,
\end{equation}
where $R$ is the Ricci scalar, $L$ is the AdS curvature scale, $\lambda$ is the tension of the brane and $h$ is the determinant of the induced metric on the defect. For brane tension in the range $0<\lambda<\frac{1}{4 \pi  \Gn L}$, the gravitational equations of motion admit stable solutions which include a thin AdS$_2$ brane. These solutions preserve the symmetries expected from a dual CFT with a conformal defect. The full solution reads
\begin{equation}\label{DCFTsol1}
ds^2 = L^2\left(d \bar y^2 + \cosh^2(|\bar y|-y^*)(-\cosh^2 r  dt^2+ dr^2)\right),
\end{equation}
with
\begin{equation}\label{eq:defparam}
\tanh y^* = 4 \pi  \Gn  L \lambda \,,
\end{equation}
and the brane is situated at $\bar y=0$, where one finds a discontinuity of the extrinsic curvature. These coordinates correspond to a foliation in terms of AdS$_2$ slices with coordinates $(r,t)$ on each slice.

\begin{figure}
	\centering
	\includegraphics[scale=0.8]{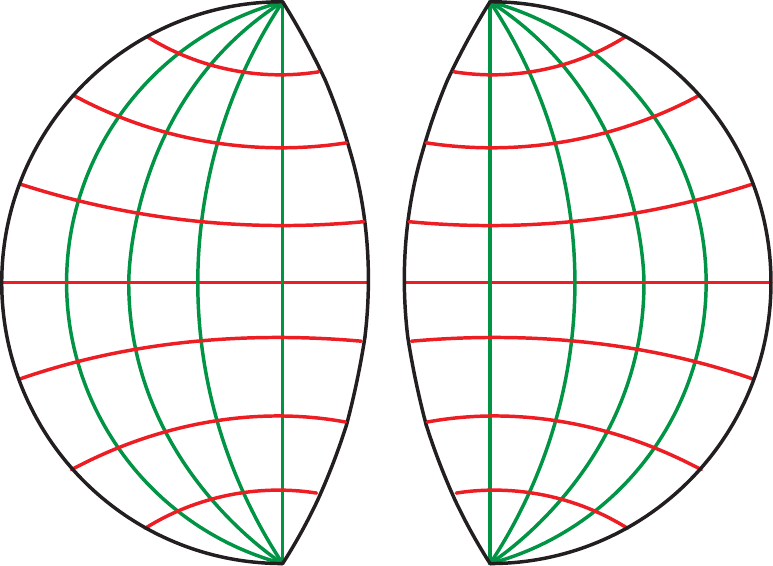}
	\caption{Constant time slices of the two AdS patches on the two sides of the defect, corresponding to the metric \eqref{DCFTsol2},  glued together at the location of the defect along $y=\pm y^*$ curves. Lines of constant $r$ are indicated in red and lines of constant $y$ are indicated in green.} \label{fig:patchesYR}
\end{figure}

The solution \eqref{DCFTsol1} can also be seen as two (slightly larger than half) patches of vacuum AdS$_3$, glued together at the location of the defect.\footnote{To be more precise, the coordinate system \eqref{DCFTsol1}, with $r>0$ only covers half of each patch. We will later translate our expressions to global coordinates where the full patches are covered.} This is most simply seen by redefining $y=\bar y- y^*$ for $\bar y>0$ ($-y^*<y<\infty$) and $y=\bar y +y^*$ for $\bar y<0$ ($-\infty<y<y^*$). Of course this coordinate system has a discontinuity at the position of the brane. The metric is then given on each patch by,
\begin{align}\label{DCFTsol2}
ds^2=  L^2\left(d y^2 + \cosh^2 y(-\cosh^2 r dt^2+ dr^2)\right),
\end{align}
where $-y^*<y<\infty$ on one side of the defect and $-\infty<y<y^*$ on the other. The ranges of the coordinates indicate that we have two patches of AdS$_3$ bounded by curves of constant $y=\mp y^*$. A cross section of the two patches, as well as their constant $y$ and $r$ slices are depicted in figure \ref{fig:patchesYR}. In this coordinate system, the boundary of AdS$_3$ is located at $y=\infty$ or $r=\infty$, constant $r$ lines are geodesics, constant $y$ curves are perpendicular to $r=0$ and constant $r$ curves are perpendicular to the boundary.

The $(t,y,r)$ coordinates can be related to the usual global AdS$_3$ coordinates $(t,\rho,\theta)$ in the following way,
\begin{equation}\label{eq:coordtrans}
\cosh y \cosh r = \cosh \rho;~~~~~~~\sinh y = \sinh \rho \sin \theta,
\end{equation}
and the metric becomes
\begin{equation}\label{metric2sol}
ds^2 = L^2 \left(-\cosh \rho^2 dt^2 + d\rho^2+\sinh ^2\rho d\theta^2 \right).
\end{equation}

Another useful set of coordinates maps the global AdS$_3$ coordinates to a circle of finite radius, it reads
\begin{equation}\label{eq:coordtrans22}
\tan \phi = \sh \rho,
\end{equation}
which leads to the following metric,
\begin{equation}\label{eq:metric.t.phi.theta}
ds^2 = {L^2 \over \cos^2\phi} \left( -dt^2 + d\phi^2 +\sin^2 \phi \, d\theta^2\right).
\end{equation}
In this way, we obtain a third coordinate system $(t,\phi,\theta)$ where the constant time slices are Poincar\'e disks with $\phi\in[0,\pi/2]$ playing the role of a radial coordinate and $\theta$ an angular coordinate on the disk. Note that constant $\theta$ curves are geodesics. The $(t,y,r)$ coordinates are related to the $(t,\phi,\theta)$ coordinates according to
\begin{equation}\label{eq:coorconvert2}
\tanh r = \sin\phi \cos\theta ;~~~~~~~\sinh y = \tan\phi \sin \theta,
\end{equation}
and these coordinates cover half the space.

Our previous coordinate systems \eqref{DCFTsol2}, \eqref{metric2sol} and \eqref{eq:metric.t.phi.theta} were dimensionless and so the curvature of the boundary will naturally be set by the AdS scale $L$.
In order to separate the radius of curvature of the boundary from the AdS curvature scale (see, e.g., \cite{Carmi:2017jqz}) we  rescale the time coordinate by a new length scale $\LB$ \cite{Comments}
\begin{equation}\label{taudef}
\tau=\LB \, t,
\end{equation}
which will set the curvature of the spatial geometry of the boundary. This leads to the metric
\begin{equation}\label{eq:metric.tau.phi.theta}
ds^2 = {L^2 \over \cos^2\phi} \left( -{d\tau^2\over \LB^2} + d\phi^2 +\sin^2 \phi d\theta^2\right)\,
\end{equation}
where the boundary time is now given by $\tau$.

\subsection{Fefferman-Graham Expansion and the Cutoff Surface}\label{subsec:cutoff}
The gravitational observables that come into play in the two holographic complexity conjectures \eqref{eq:CVformula}-\eqref{eq:CAformula} yield  divergent results and need to be regularized. The standard procedure, used also in previous studies of the holographic complexity, is to introduce a cutoff of constant $z=\delta$ in a Fefferman-Graham (FG) expansion of the relevant metric (see, e.g., \cite{Formation,Comments}). In the case of vacuum AdS$_3$, ignoring the defect, one needs to bring the metric \eqref{eq:metric.tau.phi.theta} to the form
\begin{equation}\label{metricFG2}
ds^2 = \frac{L^2}{z^2} \left( dz^2+g_{ij}(x,z) dx^i dx^j \right),
\end{equation}
where the boundary is situated at $z=0$ and $g_{ij}(x,z)$ can be expanded in a power series in $z$ where $g_{ij}(x,z=0)$ is the boundary metric.
This can be achieved by the following coordinate transformation
\begin{equation}\label{zzzz}
z=2 \LB \, \frac{\cos(\phi/2)-\sin(\phi/2)}{\cos(\phi/2)+\sin(\phi/2)},
\end{equation}
and of course, scaling the metric in the asymptotic region by $z^2/L^2$ then yields the boundary metric
\begin{equation}
ds_{bdy}^2=-d\tau^2 + \LB^2 d\theta^2. 
\end{equation}
The FG cutoff $z=\delta$ is then expressed using eq.~\eqref{zzzz} as
\begin{equation}\label{eq:cutoff}
\phi = \pi/2 -\hat\delta+\mathcal{O}(\hat \delta^3), \qquad \text{or} \qquad \cosh y \, \cosh r=\frac{1}{\sin \hat\delta}+\mathcal{O}(\hat \delta)
, \qquad \hat{\delta}\equiv {\delta\over \LB}.
\end{equation}

In the presence of the defect, however, it was shown in \cite{Papadimitriou:2004rz} that the Fefferman-Graham expansion breaks down near the defect and fails to cover a bulk wedge-shaped region originating from the defect. Different solutions to this problem have been proposed in the literature,
see the discussion in \cite{Gutperle:2016gfe} and references therein. In particular, one suggestion is to use two different cutoffs, one for the region near the defect and another one away from the defect; in the defect region, the cutoff is expressed in term of the FG coordinates of an AdS$_2$ slicing of the geometry (in our coordinates \eqref{DCFTsol2} these are the slices of constant $y$).
We adopt this suggestion for regularizing the complexity in the defect region, but continue to use the standard FG cutoff away from the defect. Moreover, we choose the two cutoffs in such a way that the cutoff surface is smooth.\footnote{Although we do not have a strong justification, it seems natural to require smoothness of the cutoff surface; moreover this avoids some problems that would arise in the CA computation where a lack of smoothness would introduce additional joints, see footnote \ref{strangeJoints}. We would like to thank Rob Myers for suggesting this choice of cutoff.}

\begin{figure}
	\centering
	\includegraphics[scale=1.0]{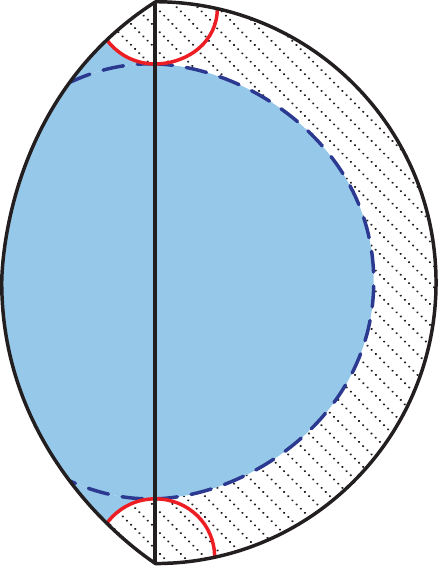}
	\caption{Extension of the cutoff surface in the region of the defect following lines of constant $r$. This generates a cutoff surface which is perpendicular to the defect and connects smoothly the two sides. We have indicated in light blue the region inside the cutoff surface.} \label{fig:Cutoff}
\end{figure}

Note that naively extending the cutoff surface of constant radius $\phi$ in eq.~\eqref{eq:cutoff} up to the defect would give a surface that does not  match smoothly across the defect. One way to see this is to check that the FG cutoff surface is not perpendicular to the line of constant $y=y^*$, where it crosses the defect. The proposed  extension of the cutoff surface in the region of the defect ($y<0$)  is
given by constant $r$ curves on each side, see figure \ref{fig:Cutoff}. Explicitly the constant $r$ extension of the cutoff surface can be expressed using eq.~\eqref{eq:coorconvert2} and it reads
\begin{equation}\label{eq:cutoff2}
\tanh r = \sin\phi \cos\theta = \cos\hat{\delta} \,.
\end{equation}
The two parts of the cutoff surface \eqref{eq:cutoff} and \eqref{eq:cutoff2} are smoothly connected at $y=0$.

\subsection{Wheeler-DeWitt Patch in Defect AdS$_3$}\label{sec:WDWpatch}
\begin{figure}
	\centering
	\includegraphics[scale=0.42]{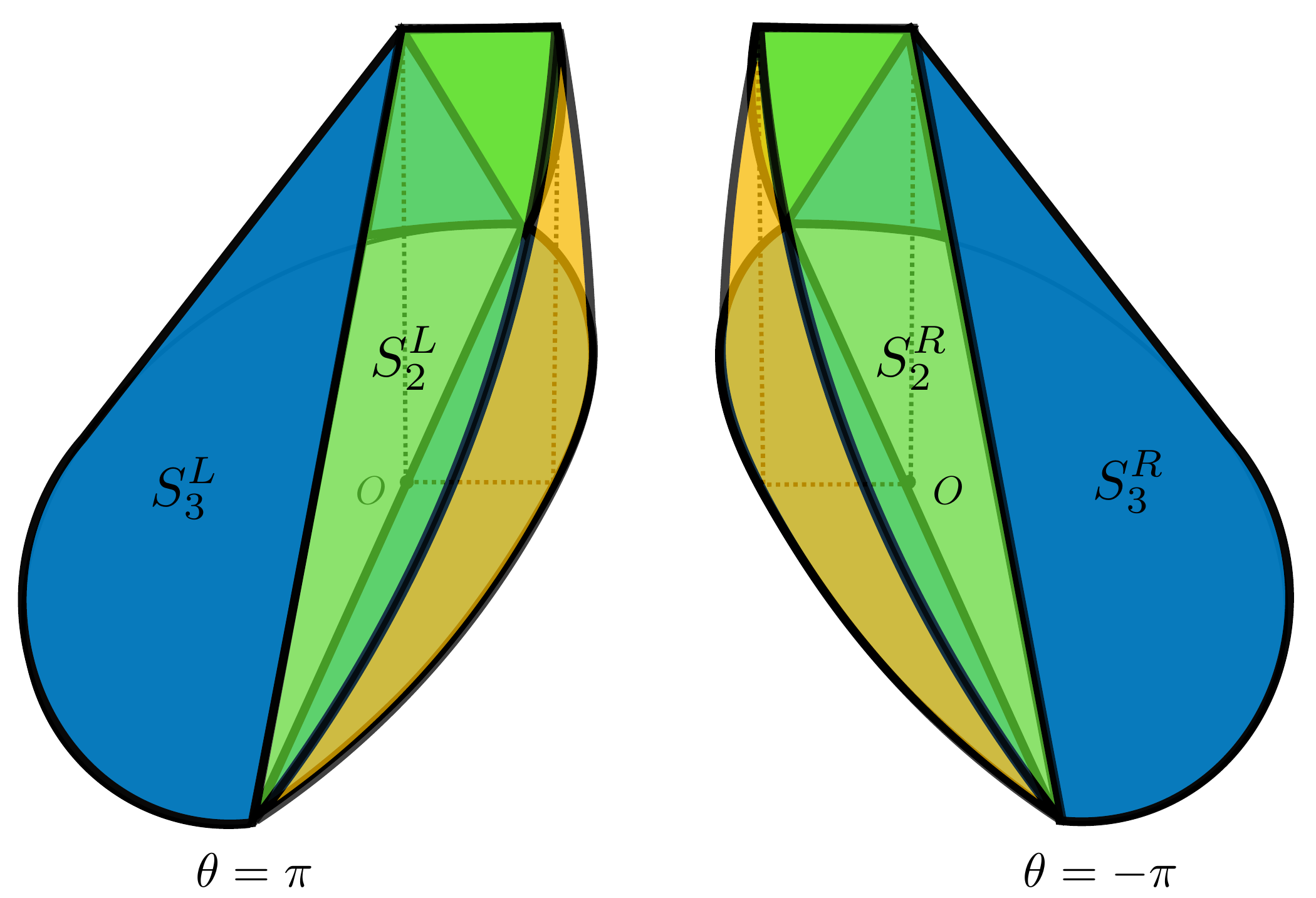}
	\caption{Illustration of the (future half of the) WDW patch in defect AdS$_3$. $S_3^L$ and $S_3^R$ are the two half cones, already present for the case of vacuum AdS$_3$. $S_2^L$ and $S_2^R$ are the additional boundaries of the WDW patch in the defect region, fixed by parts of the lightcones generated from the points $\theta=0,~\pm\pi$ on the boundary ($\phi=\pi/2$). Those null surfaces are smoothly connected across the defect and they terminate along a ridge at the top of the WDW patch. The yellow surfaces correspond to the defect brane, where the left and right patches are glued together.}
	\label{fig:WDW}
\end{figure}

The Wheeler-DeWitt (WDW) patch is defined as the union of all spacelike surfaces anchored at the boundary time slice where the state is defined. A practical way to obtain its shape is to identify the parts of space which are not contained within the lightcones generated from any of the points on the given boundary time slice. Without loss of generality we choose this time slice to be $t=0$. In the case of pure AdS$_3$ (without the defect), the WDW patch takes the form of a cone generated from the relevant time slice on the boundary, bounded by light sheets (see, e.g., the left panel of figure 2 in \cite{Formation}). In the defect geometry however, the WDW patch will be bounded by additional surfaces in the defect region, see figure \ref{fig:WDW}. Those surfaces correspond to the lightcones generated from the points at the intersection of the boundary and the defect on the $t=0$ time slice, namely $\theta=0$ and $\phi=\pi/2$ or $\theta=\pm\pi$ and $\phi=\pi/2$. To understand the shape of these extra boundaries of the WDW patch, we need to obtain the relevant surfaces in the defect region by explicitly analyzing the lightcone generated from a given point on the boundary, e.g., $\theta=0$.\footnote{The result for $\theta=\pm\pi$ is easily obtained using symmetry arguments.} As we demonstrate below, the lightcone takes a very simple form in the $(t,y,r)$ coordinate system.

\begin{figure}
	\centering
	\includegraphics[scale=0.45]{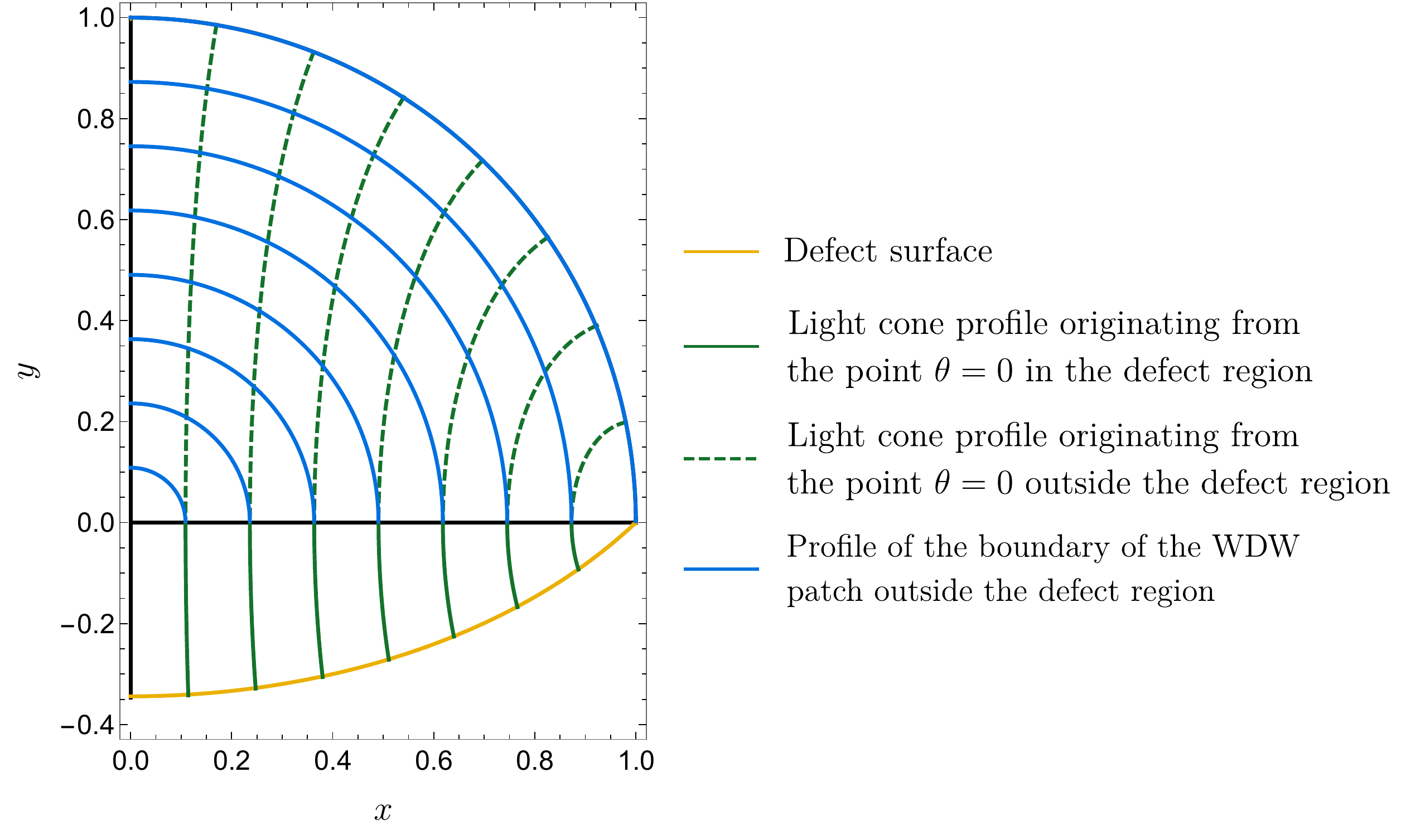}
	\caption{Cross section of the boundary of the WDW patch for different times $t$ denoted by the green and blue lines inside and outside the defect region respectively. In the defect region the boundary of the WDW patch is fixed by the light cone emanating from the boundary at $\theta=0$, indicated by solid green curves, and  it  meets the lightcone surface coming from $\theta=\pi$ along a ridge at $\theta=-\pi/2$ (and $t=\pi/2$). The rest of the boundary of the WDW patch is the conical region fixed by straight infalling light rays coming from different boundary points along lines of constant $\theta$ and its cross section for different times $t$ is indicated by the blue circular arcs.  The plot corresponds to a  defect parameter of $y^*=0.6$, see eq.~\eqref{eq:defparam}, and is presented using the $x$ and $y$ coordinates defined in eq.~\eqref{eqapp:xycoords}.}
	\label{fig:ProfileWDWPatch}
\end{figure}

We will study the null geodesics starting from the boundary point $t=0$, $r=\infty$ in the metric \eqref{DCFTsol2}. Since the $(y,r)$ coordinate system is singular at $r=\infty$, one of the initial conditions is replaced by a  regularity condition at this point, which as we show below amounts to having the geodesic follow an initial angular orientation along some $y=y_0$, with $\dot y=0$, where the derivative is taken with respect to some parameter $\sigma$ along the null geodesics. In appendix \ref{app:nullGeoApp} we derive the same geodesics directly in global coordinates as a consistency check. With the change of variables $\tanh (r(\sigma))=R(\sigma)$ and $\tanh (y(\sigma))=Y(\sigma)$ and a choice of parametrization $\sigma=t$ we obtain the following equations of motion by minimizing the line element
\begin{equation}\label{eq:null.geo.eqs}
\ddot R(t)= -R(t),
\qquad
\ddot Y= Y \left[\frac{\dot R^2}{\left(R^2-1\right)^2}+\frac{1}{R^2-1}\right]
-\frac{2 R  \dot R \dot Y}{R^2-1}
,
\end{equation}
and the requirement that the geodesics are null, namely the vanishing line element, reads
\begin{equation}\label{eq.null.const}
\dot Y^2=
(Y^2-1) \left[\frac{\dot R^2}{\left(R^2-1\right)^2}+\frac{1}{R^2-1}\right] .
\end{equation}
Eq.~\eqref{eq:null.geo.eqs} is solved by
\begin{equation}\label{eq.null.ans1}
R(t) = c_1 \cos t + c_2 \sin t .
\end{equation}
The boundary condition $R(t=0)=1$ fixes $c_1=1$ and substituting eq.~\eqref{eq.null.ans1} into eq.~\eqref{eq.null.const} we find that $\dot y$ diverges at $t=0$ unless $c_2=0$. This constraint is analogous to setting to zero the angular momentum of a geodesic passing through the radial origin of a polar coordinate system. The null equation is then solved by $Y(t)=Y_0$ where $Y_0$ is a constant, and this also solves the second equation of \eqref{eq:null.geo.eqs}. Reverting the change of variables we finally obtain
\begin{equation}\label{lightconeryt}
\tanh  r  = \cos t, \qquad y=y_0.
\end{equation}
This means that the null geodesics are following lines of constant $y$ while $r$ and $t$ are changing. Because of this fact it is very natural to work with $y,r$ coordinates in the defect region, while we will keep working with the $\phi,\theta$ coordinates outside the defect region. A cross section of this surface for different values of $t$ is depicted by the green slices in figure \ref{fig:ProfileWDWPatch}. We see that the constant time slices on our null cone straighten up as we go deeper into the bulk and they finally follow the constant angular surface of $\theta=-\pi/2$. We conclude that the two new boundaries of the WDW patch, see the green surfaces in figure \ref{fig:WDW}, meet along a ridge at $\theta=-\pi/2$ and $t=\pi/2$. For $y>0$ the WDW patch is fixed by the light rays which come from other boundary points, and its constant $t$ profile is represented in figure \ref{fig:ProfileWDWPatch} as blue curves which correspond to the blue conical surface in figure \ref{fig:WDW}. This surface is the same as the boundary of the WDW patch in the absence of the defect.

\section{Holographic Complexity with a Defect}\label{sec:CVCA}
With the geometric understanding developed in the previous section, we are now ready to investigate the predictions of the two holographic proposals  (CA and CV) for the complexity of the DCFT ground state in our holographic defect toy model.

\subsection{CV Conjecture}
\begin{figure}
	\centering
		\centering \includegraphics[scale=0.4]{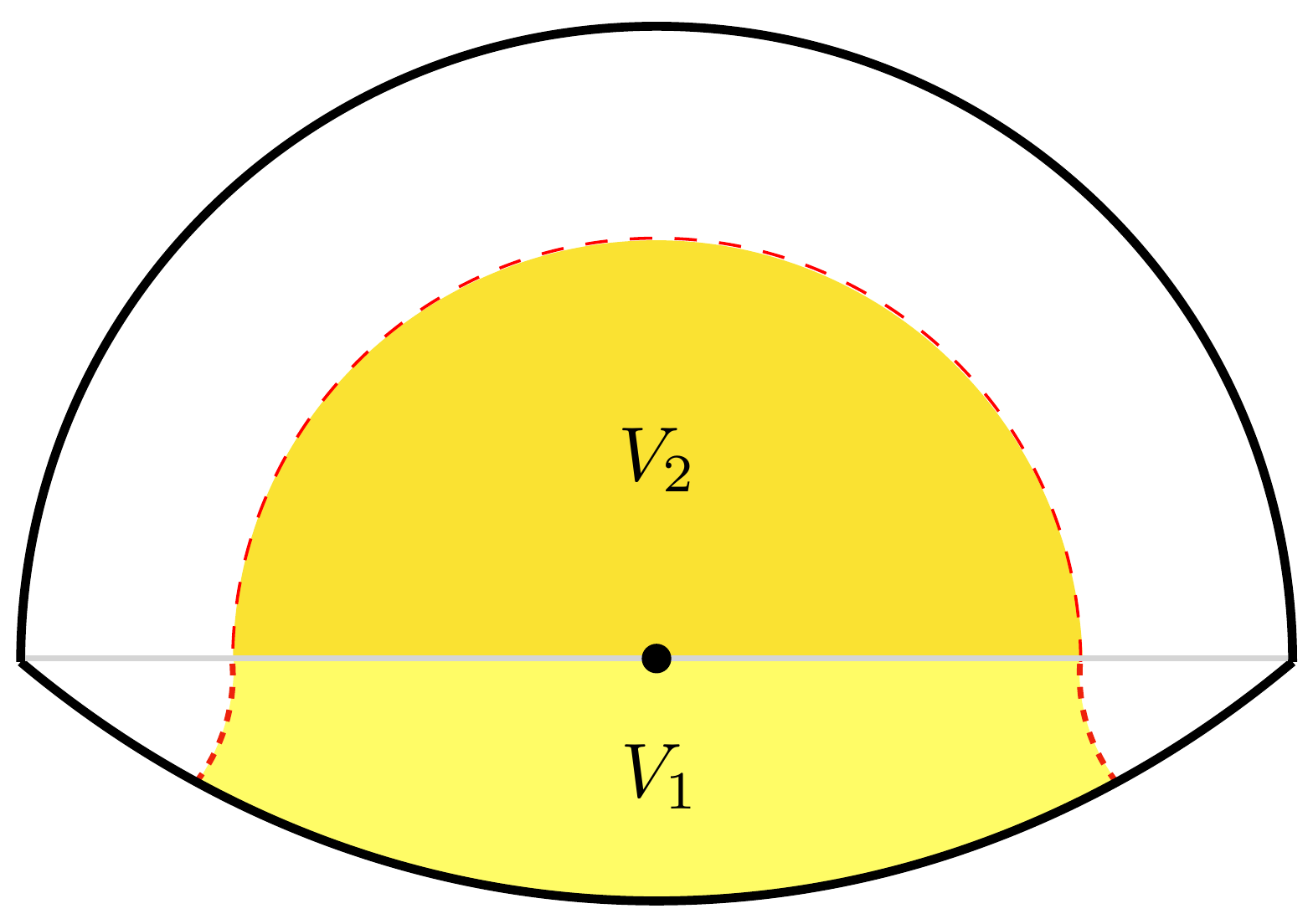}
        \caption{Division of the constant time slice inside the cutoff surface to two different portions which we use in evaluating the volume integrals for the CV conjecture. $V_1$ is the volume in the defect region and $V_2$ is the volume outside the defect region.} \label{fig:CVRegions1}
\end{figure}

We start with the CV conjecture \eqref{eq:CVformula}. In this case we have to evaluate the volume of a constant time slice in the presence of the defect. Since the two sides of the defect are identical, we focus on the region $-y^*<y<\infty$ below. We will eventually multiply the final result by two in order to account for the two sides of the defect.
The defect is located at constant $y=-y^*$ which according to eqs.~\eqref{eq:coordtrans} and \eqref{eq:coordtrans22} corresponds to
\begin{equation}\label{eq:defb}
\tan\phi \sin \theta =-\sh y^*.
\end{equation}
Since the volume is divergent we will use the cutoff surface in eqs.~\eqref{eq:cutoff} and \eqref{eq:cutoff2}, see figure \ref{fig:Cutoff}. We divide our volume to two parts $V_1$ and $V_2$ as indicated in
figure~\ref{fig:CVRegions1}. Integrating the volume element, given by the square root of the induced metric, in each of these regions yields
\begin{equation}\label{eq:CVres123}
\begin{split}
V_1= &\,  2L^2  \int_{-y^*}^{0} dy \cosh y \int_0^{\tan^{-1}(\cos \hat \delta)} dr = 2L^2 \sinh y^*  \ln \left(\frac{2}{\hat \delta}\right)\,,
\\
V_2= &\, L^2 \int_0^\pi d\theta\int_0^{\pi/2-\hat{\delta}} d\phi\, {\sin\phi \over \cos^2\phi}=L^2\left(\frac{\pi }{\hat \delta}-\pi\right)\,,
\end{split}
\end{equation}
up to terms of order $\hat\delta$.
Summing everything up and using eq.~\eqref{eq:CVformula} we obtain the following result for the complexity using the CV proposal
\begin{align}\label{eq:CVresult}
\mC_V= \frac{2}{\Gn L} \left(V_1+V_2\right)
=\frac{4 c_{{}_T}}{3}\left({\pi\over \hat\delta} + 2\sinh y^* \ln\left({2 \over \hat\delta}\right) -\pi\right)
\end{align}
where we have included an overall factor of 2 to account for the two sides of the defect and expressed the result in terms of the central charge $c_{{}_T}=3 L/(2\Gn)$. The leading contribution is the same as in the case without the defect and it follows a volume law (recall from eq.~\eqref{eq:cutoff} that $\hat \delta = \delta/L_{\mathcal{B}}$). We see that the contribution introduced by the defect includes a logarithmic UV divergence with a coefficient which is proportional to $\sinh y^*$ where $y^*$ is related to the tension of the brane according to eq.~\eqref{eq:defparam}. For a brane with small tension for instance we will have a linear relation $\sinh y^*\sim y^*\sim\lambda$. On the CFT side we expect the relevant parameter to encode properties of the defect CFT. Of course, this result is larger than in the absence of the defect since due to the defect, the space was extended and so the volume has increased. We also note that the result is proportional to the central charge (equivalently, the number of degrees of freedom in the system). We will compare these results to those of simple CFT models with defects in section \ref{sec:QFT}.

\subsection{CA Conjecture}\label{subsec:CAfull}
\begin{figure}
	\centering
	\begin{subfigure}[b]{0.3\textwidth} 
		\centering \includegraphics[width=\textwidth]{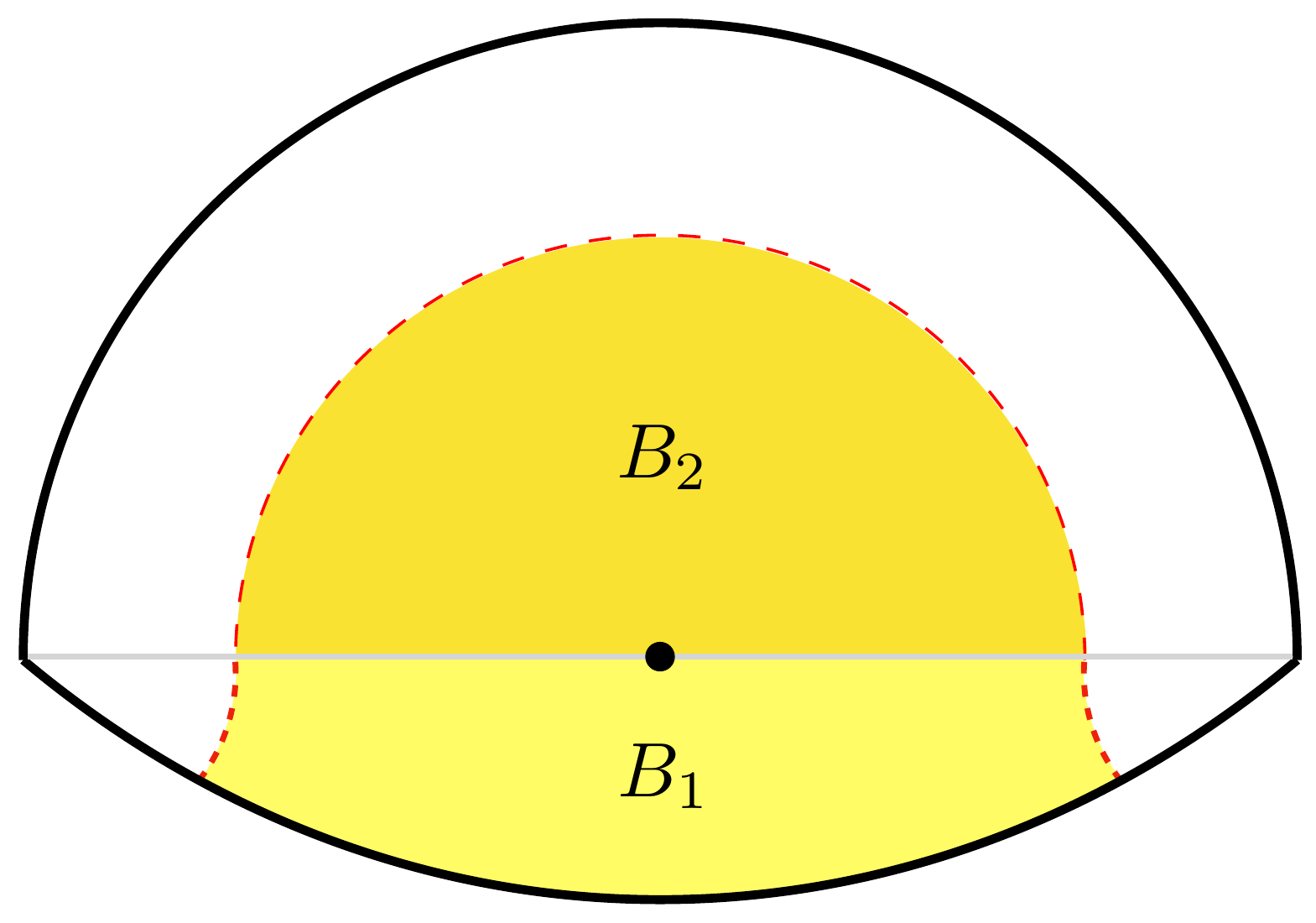}
		\caption{Bulk contributions}\label{fig:CA_vol}
	\end{subfigure}
	~ 
	\begin{subfigure}[b]{0.3\textwidth}
		\centering \includegraphics[width=\textwidth]{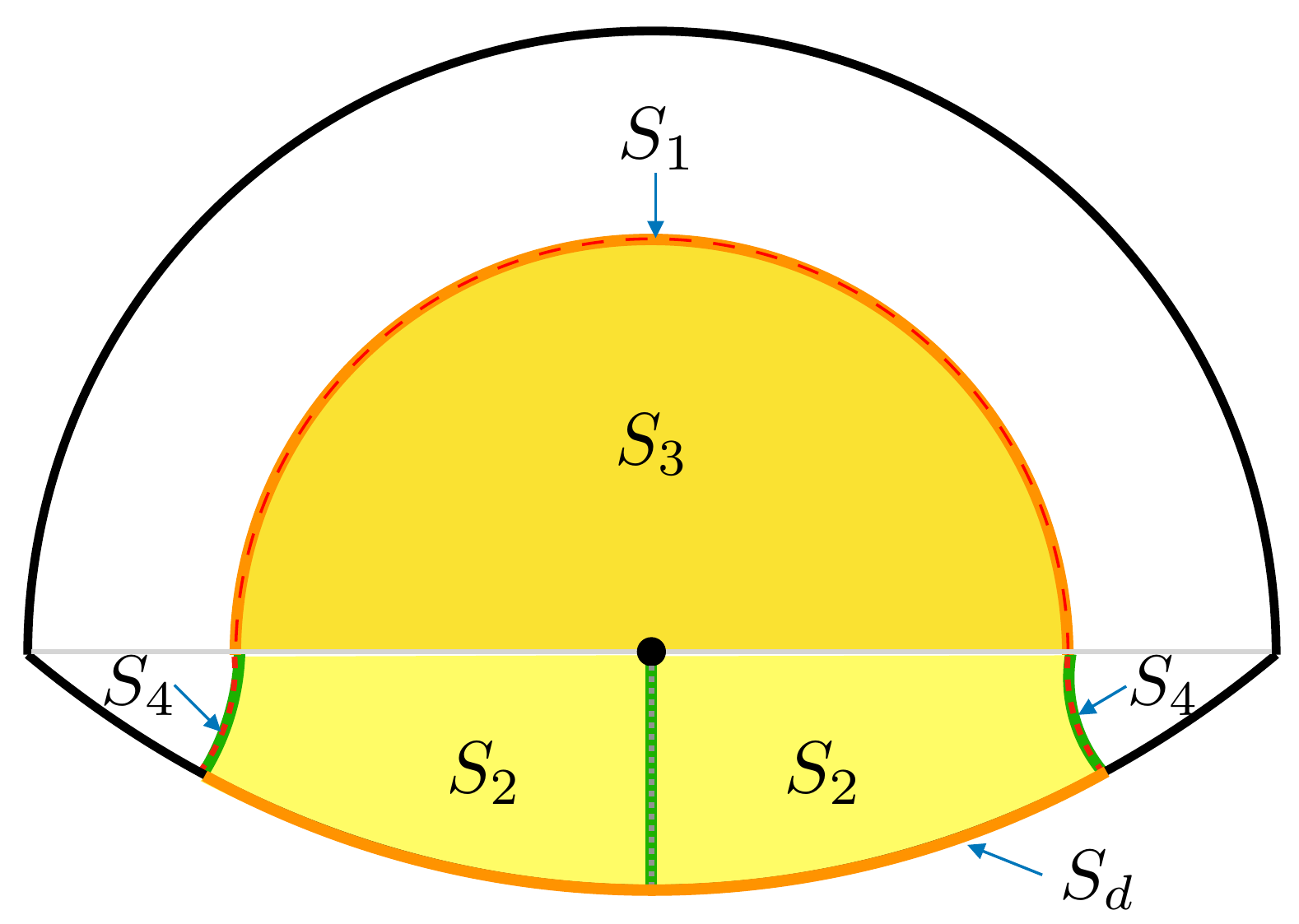}
		\caption{Surface contributions}\label{fig:CA_surf}
	\end{subfigure}
	~
	\begin{subfigure}[b]{0.3\textwidth}
		\centering \includegraphics[width=\textwidth]{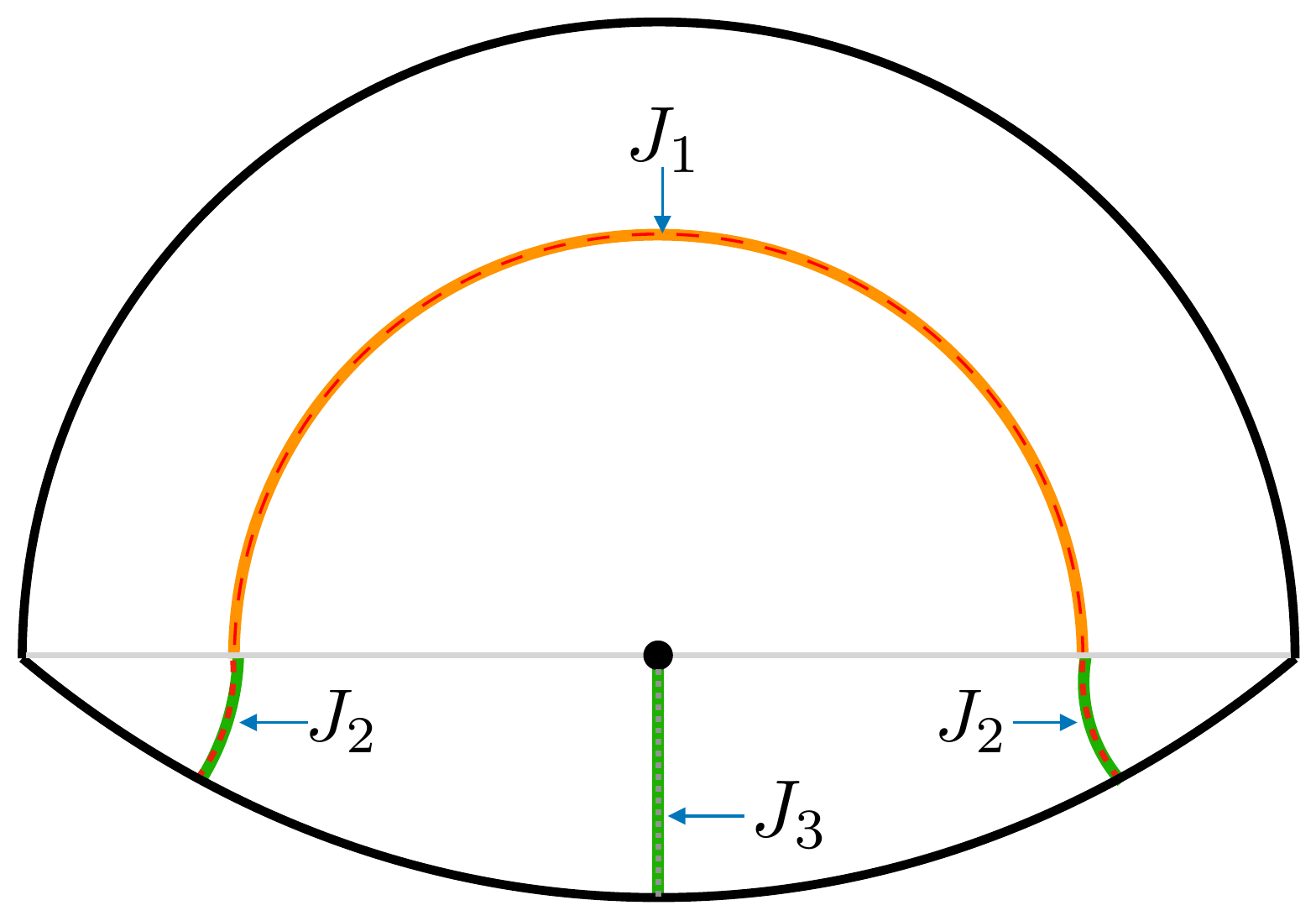}
		\caption{Joint contributions}\label{fig:CA_joint}
	\end{subfigure}
	\caption{Different contributions for the CA conjecture projected onto a constant time slice. The bulk contributions consist of those above and below the indicated regions ($B_1$ and $B_2$). The surface contributions include $S_1$ and $S_4$ which are due to the cutoff surface and $S_2$ and $S_3$ which are due to the null boundaries of the WDW patch. $S_d$ stands for the defect contribution. The joint contributions consist of $J_1$ and $J_2$ which stand for the joints at the intersection of the cutoff surface and the null boundaries of the WDW patch and of $J_3$ which stands for the joint at the ridge at the top of the WDW patch, see also figure \ref{fig:WDW}.}
	\label{fig:CA}
\end{figure}

Next we evaluate the complexity using the CA conjecture \eqref{eq:CAformula}, which states that, up to an overall numerical coefficient, the complexity is given by the gravitational action of the WDW patch. The gravitational action consists of a number of different contributions including bulk (codimension-0), boundary (codimension-1) and joint (codimension-2)
terms.\footnote{We also encounter caustics, e.g., at the tip of the blue cone in figure \ref{fig:WDW}. The contribution of the tip can effectively be calculated by regulating it using a cutoff surface at constant $t=\pi/2-\epsilon$. In this way we are able to demonstrate that this caustic does not make an additional contribution to the gravitational action by smoothly taking the limit $\epsilon \rightarrow 0$. We are not aware of an explicit prescription for such contributions in the literature, but we would like to point out that it is hard to come up with an action for such point-like elements which is consistent with dimensional analysis since (before dividing by $\Gn$) it should have mass dimension $-1$.}
The relevant contributions involving null joints have recently been analyzed in \cite{Lehner:2016vdi}, which we will follow in our calculation below, and other relevant boundary and joint contributions were previously explored  in \cite{Gibbons:1976ue,York:1972sj,Parattu:2015gga,Hayward1993,Brill:1994mb}.
For the current setup the relevant contributions in the gravitational action read
\begin{equation}\label{eq:CAaction}
\begin{split}
I= &\, \frac{1}{16 \pi \Gn} \int_\mathcal{M}d^{3}x \sqrt{-g}\left(R-2\Lambda\right)
 +\frac{\epsilon_{{}_K}}{8 \pi \Gn}\int_{\mathcal{B}_{t/s}} d^2x \sqrt{|h|} K
 \\
 &\, + \frac{\epsilon_\kappa}{8\pi\Gn}\int_{\mathcal{B}_n} d\lambda\, d\theta \sqrt{\gamma} \,
 \kappa
- {1\over 8\pi\Gn}\int_{\mathcal{B}_n} d\lambda\, d\theta \sqrt{\gamma} \,
  \Theta \ln (\ell_{ct} | \Theta|)
 \\
 &\,
 + \frac{\epsilon_{\mathfrak{a}}}{8\pi\Gn}\int_\Sigma dx \sqrt{\gamma} \, \mathfrak{a} -\lambda \int_{D\, \cap\, WDW} d^2 x \sqrt {-h}\, .
\end{split}
\end{equation}
The various contributions are: the bulk Einstein-Hilbert action with negative cosmological constant; the Gibbons-Hawking-York (GHY) extrinsic curvature term for timelike/spacelike boundaries;\footnote{We will only need to evaluate the GHY contribution for timelike surfaces. In this case $\epsilon_{{}_K}=1$ and the normal vector $s^\mu$ should be oriented away from the volume of interest. We evaluate the extrinsic curvature according to $K_{ab} = e^\mu_a e^\nu_b \nabla_\mu s_\nu$ and its trace is given by $K = h^{ab} K_{ab}$ where the vielbeins are defined as $e^\mu_a = \partial_a x^\mu$, the induced metric is given by $h_{ab}=g_{\mu\nu}e^\mu_a e^\nu_b$ and the indexes $a,b$ label coordinates inside the surface.} the null boundary contribution given in terms of $\kappa$ which measures how far is the parameter $\lambda$ from providing an affine parametrization of the null generators of the null surface\footnote{$\kappa$ is defined according to $k^\mu \nabla_\mu k_\nu = \, \kappa k_\nu$, where $k^\mu=d x^\mu/d \lambda$ is the future oriented null normal vector and $\lambda$ is a parameter along the null generators increasing toward the future. As noted in reference \cite{Chapman:2018dem}, the $\kappa$ term in references \cite{Lehner:2016vdi} and \cite{Carmi:2016wjl} had a sign mistake which we corrected for in eq.~\eqref{eq:CAaction}. $\epsilon_\kappa=\pm 1$ if the volume of interest lies to the future (past) of the boundary segment.\label{kappafoot}} and a counterterm added in order to ensure parametrization invariance given in terms of the null expansion $\Theta$;\footnote{The expansion parameter is defined according to $\Theta = \partial_\lambda \ln \sqrt{\gamma}$ where $\gamma$ is the (one dimensional) metric on the null surface. The addition of this counterterm was recently pointed out to be an essential ingredient of the CA conjecture in refs.~\cite{Chapman:2018dem,Chapman:2018lsv}.} contributions from spacelike joints involving null surfaces given in terms of $\mathfrak{a}$;\footnote{$\mathfrak{a}$ is given by $\mathfrak{a}=\ln |s \cdot k|$ for the case of the intersection between a timelike and a null boundary with normal vectors $s$ and $k$ respectively,  and by $\mathfrak{a}=\ln |k_1 \cdot k_2/2|$ for the intersection between two null boundaries with normal vectors $k_1$ and $k_2$. The sign $\epsilon_{\mathfrak{a}}=-1$ if the volume of interest lies to the future (past) of the null segment and the joint lies to the future (past) of the segment and $\epsilon_{\mathfrak{a}}=1$ otherwise. For more details see appendix C of \cite{Lehner:2016vdi}.} and the new contribution due to the gravitational action for the defect itself in the region enclosed inside the WDW patch. We excluded from the action \eqref{eq:CAaction} other joint contributions which did not enter our calculations, see appendix C of \cite{Lehner:2016vdi}. In the following subsections we evaluate all these contributions and finally sum them up to produce a result for the complexity at the end of the section. The shape of the WDW patch has already been described in subsection \ref{sec:WDWpatch}, see figure \ref{fig:WDW}, and the various contributions are depicted and enumerated in figures \ref{fig:CA} and \ref{fig:Onewedge}.

\subsubsection{Bulk Contributions}
We start by evaluating the bulk Einstein-Hilbert and cosmological constant contributions.
The Ricci scalar is the same as for the case of vacuum AdS$_3$ everywhere except at the position of the brane where it has an extra delta function. The effect of this additional delta function integrated over the infinitesimal thickness of the brane will be dealt with later on, together with the brane action contribution in subsection \ref{subsec:defdef}.
For the case of vacuum AdS$_3$ we have $R=-6/L^2$ and $\Lambda=-1/L^2$ and therefore
\begin{equation}
I_{\text{bulk}}\equiv\frac{1}{16 \pi \Gn} \int_\mathcal{M}d^{3}x \sqrt{-g}\left(R-2\Lambda\right)=
-{1\over 4\pi\Gn}\int_{\mathcal{M}}d^3x {\sqrt{-g}\over L^2}.
\end{equation}
This will allow us to evaluate the Einstein-Hilbert contribution everywhere except for an infinitesimally thin shell surrounding the brane. The relevant contributions can be divided into two regions $B_1$ and $B_2$ whose projections on a constant time slice are depicted in figure \ref{fig:CA_vol}. Due to the symmetries of the problem, we focus on the future part of the WDW patch on one side of the defect and eventually multiply our result by a factor of four. We start from the contribution of the region $B_1$:
\begin{equation}\label{B1t}
\begin{split}
B_1 =& \, -\frac{L}{2\pi\Gn}\int_{-y^*}^{0} dy \cosh^2 y \int_{0}^{\tanh^{-1} (\cos \hat \delta)} dr \cosh r \int_{0}^{\cos^{-1}(\tanh r)} dt\,
\\
= &\, \frac{L}{4\pi\Gn} \left[ y^* + \frac{1}{2} \sinh(2 y^*)\right] \left( \ln \hat{\delta}-1\right)\,.
\end{split}
\end{equation}
Next, we evaluate the contribution from the region $B_2$:
\begin{equation}\label{B2t}
B_2=-\frac{L}{4\pi\Gn}\int_0^{\pi} d\theta \int^{\pi/2 - \hat{\delta}}_0 d\phi \int_0^{\pi/2 -\phi} dt \frac{ \sin\phi}{\cos^3\phi}
=-\frac{L}{4\pi\Gn} \left[\frac{\pi}{\hat{\delta}}-\frac{\pi^2}{4}\right].
\end{equation}
Summing together eqs.~\eqref{B1t} and \eqref{B2t} and multiplying by a factor of four for the two sides of the defect as well as the future and past parts of the WDW patch we obtain
\begin{equation}
\begin{split}
I_{\text{bulk}}= & \,
\frac{L}{\pi\Gn} \left( - \frac{\pi}{\hat{\delta}}+ \gamma (\ln \hat{\delta}-1)
+\frac{\pi^2}{4} \right), \qquad \gamma\equiv y^* + \frac{1}{2} \sinh(2 y^*)\,,
\end{split}
\end{equation}
where the parameter $\gamma$ encodes the influence of the defect. For a brane with small
tension we have a linear relation between $\gamma$ and the tension of the brane, namely $\gamma\sim \lambda$.

\subsubsection{Boundary and Joint Contributions}\label{sec:CAsurfaces}
\begin{figure}
	\centering
		\centering \includegraphics[scale=0.6]{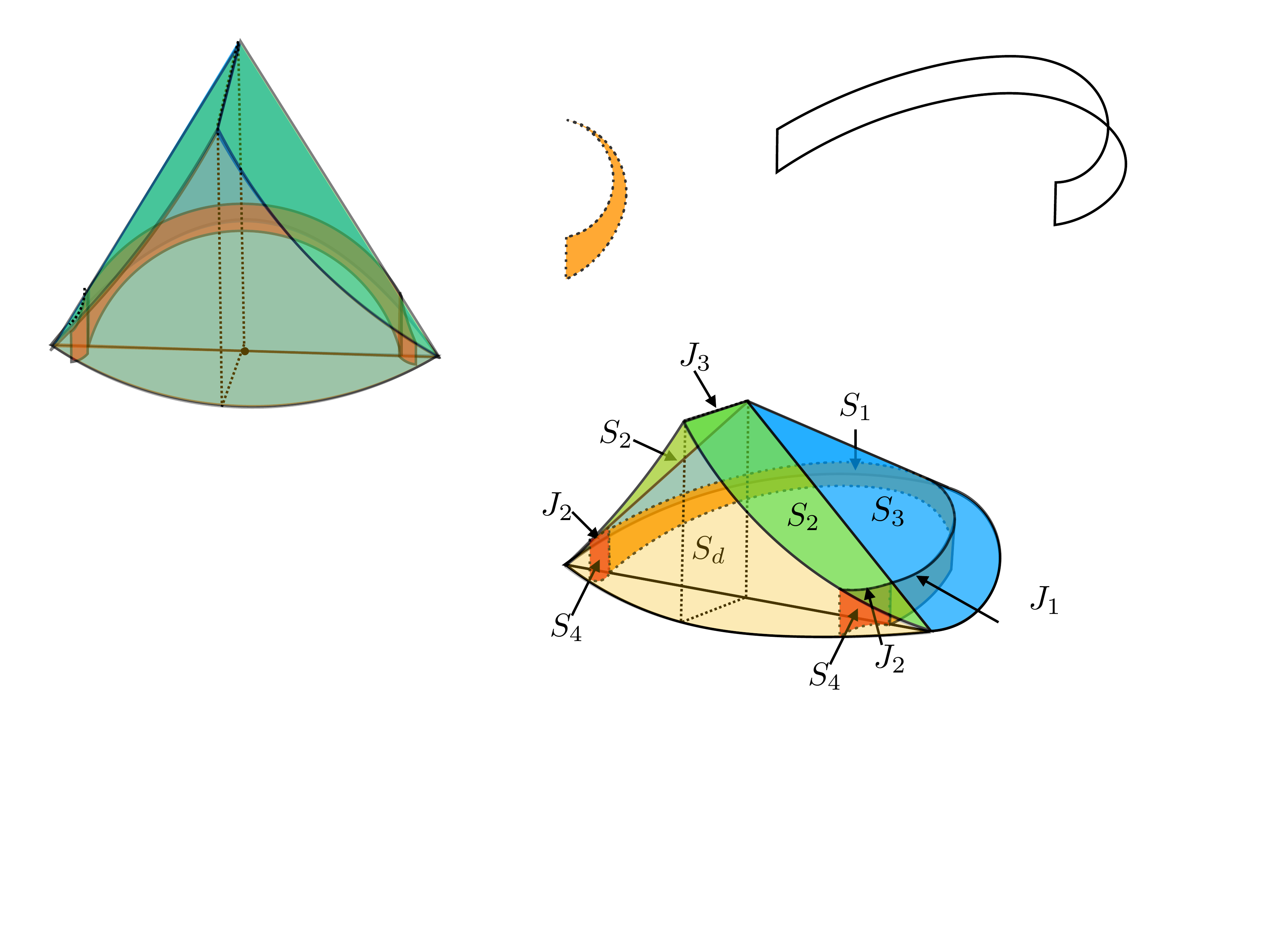}
        \caption{Various joint and surface contributions to the action of the WDW patch. We have focused on the future half of the patch on one side of the defect.} \label{fig:Onewedge}
\end{figure}

In this section we evaluate the various boundary and joint contributions to the gravitational action of the WDW patch. The different surfaces and joints which come into play in this calculation are illustrated and labeled in figure \ref{fig:Onewedge} and their projections on a constant time slice are presented in figures \ref{fig:CA_surf}-\ref{fig:CA_joint}. They consist of the half cylindrical cutoff surface outside the defect region which is labeled as $S_1$, the two additional null boundaries consisting of lightcones generated from antipodal points on the boundary of the WDW patch in the defect region, both labeled as $S_2$, the half cone outside the defect region labeled as $S_3$, and the two additional constant $r$ extensions of the cutoff surface in the defect region labeled  as $S_4$. The joint between $S_1$ and $S_3$ is labeled as $J_1$ and the one between $S_2$ and $S_4$ is labeled  as $J_2$. Finally, the joint at the ridge at the top of the WDW patch between the two $S_2$ surfaces is labeled as $J_3$. Note that $S_2$ and $S_3$ are connected smoothly, as well as the various surfaces on two different sides of the defect and we therefore do not include additional joint contributions there.

\paragraph{Contributions outside the defect region}
We start by evaluating the various contributions outside the defect region.
The half cylindrical cutoff surface $S_1$ corresponds to $\phi=\pi/2 - \hat{\delta}$ and its normal one-form and induced metric read
\begin{equation}\label{eq:S1info1}
{\bf s^{(1)}} \equiv s^{(1)}_\mu dx^\mu = \frac{L}{\sin \hat \delta} \, d\phi , \qquad
{ dh_{(1)}^2} = \frac{L^2}{\sin^2 \hat{\delta}}\left( -dt^2 + \cos ^2{\hat\delta} \, d\theta^2 \right).
\end{equation}
The extrinsic curvature reads
\begin{equation}\label{eq:S1info2}
K_{(1)}  =  \frac{1}{L} \left(\cos\hat{\delta} + \frac{1}{\cos\hat{\delta}}\right),
\end{equation}
which yields the following GHY contribution to the gravitational action
\begin{equation}\label{S1full}
S_1=
\frac{L}{8\pi\Gn} \int_0^{\hat\delta} dt \int_0^\pi d\theta \, \frac{\cos\hat{\delta}}{\sin^2\hat{\delta}} \left( \cos\hat{\delta} + \frac{1}{\cos\hat{\delta}}\right) =
\frac{L}{4\Gn\hat{\delta}}\, .
\end{equation}
The half cone $S_3$ can be parameterized by the coordinates $\lambda=t/\mathcal{N}_3$ and $\theta$ as follows\footnote{In \cite{Lehner:2016vdi} it was suggested that as a part of the prescription to evaluate the complexity we should choose a parametrization of the null generators such that $\kappa=0$ and such a parametrization is given by the choice $\lambda\propto \cot (t)$. However, since we are adding the counterterm the choice of parametrization will not modify the final result and we may proceed with $\lambda \propto t$.}
\begin{equation}
x^\mu(\lambda,\theta) \equiv (t,\phi,\theta) = (\mathcal{N}_3\lambda,\pi/2-\mathcal{N}_3\lambda ,\theta),
\end{equation}
and the normal vector to the surface reads
\begin{equation}\label{k3k3k3}
k^\mu_{(3)} = \frac{d x^\mu}{d \lambda} = \mathcal{N}_3 (1,-1,0)\,.
\end{equation}
The (one dimensional) induced metric, expansion and $\kappa$ on the surface $S_3$ are given by
\begin{equation}\label{eq:S3info}
\gamma_{\theta \theta}=L^2 \cot^2 t, \qquad \Theta =- \frac{2\mathcal{N}_3 }{\sin (2t) }, \qquad
\kappa_{(3)} = -2 \mathcal{N}_3 \cot t\,.
\end{equation}
We can now use these results to evaluate the surface contribution $S_3$, including both the $\kappa$ term and the counterterm contribution $\Theta \ln (\ell_{ct} |\Theta|)$ from eq.~\eqref{eq:CAaction}. This leads to
\begin{equation}\label{S3full12345}
S_3 = \frac{L}{8 \pi \Gn} \int^{\pi/2}_{\hat \delta} dt \int_{0}^\pi d\theta  \,  \,\frac{\ln \left(\frac{2\ell_{ct} \mathcal{N}_3}{\sin(2t)}\right)+2\cos^2  t}{\sin^2 t}
= \frac{L}{8 \Gn \hat \delta} \left( \ln \left(\frac{\ell_{ct} \mathcal{N}_3}{\hat \delta}\right)+ 1\right)\,.
\end{equation}
The joint $J_1$, where the half-cone intersects with the cylindrical cutoff surface, is given as
\begin{align}\label{J1fullfull}
J_1 =
-\frac{L}{8\pi\Gn} \int_0^\pi d\theta\, \cot(\hat\delta) \ln \left(\frac{\mathcal{N}_3 L}{ \sin \hat\delta}\right)  =
-\frac{ L}{8\Gn\hat{\delta}} \ln\left( \frac{\mathcal{N}_3 L}{ \hat\delta} \right).
\end{align}
Combining all these contributions for the surfaces and joints outside the defect region together and multiplying by a factor of four for the future and past parts of the WDW patch as well as the two sides of the defect we obtain
\begin{equation}
I_{\text{sj,out}}=4(S_1+S_3+J_1) = \frac{ L}{2\Gn\hat{\delta}} \left( \ln (\ell_{ct}/L )+ 3\right).
\end{equation}
We see that the parametrization choice $\mathcal{N}_3$ canceled out as expected due to the addition of the counterterm.

\paragraph{Contributions inside the defect region} Here we focus on the various surfaces and joints inside the defect region, namely $S_2$, $S_4$, $J_2$ and $J_3$, see figures \ref{fig:CA}-\ref{fig:Onewedge}.
The constant $r$ cutoff extension in the defect region corresponds to the constraint $\tanh r = \cos\hat{\delta}$, see eq.~\eqref{eq:cutoff2}, and its normal one-form and induced metric read
\begin{equation}
\begin{split}
{\bf s^{(4)}} &\equiv s^{(4)}_\mu dx^\mu  = L \cosh y \, dr ,\\
{ d h_{(4)}^2} &=  L^2 \left( dy^2- \frac{\cosh^2 y}{\sin^2 \hat \delta} \, dt^2 \right)\,.
 \end{split}
\end{equation}
The extrinsic curvature reads
\begin{equation}
K_{(4)}  =  \frac{\cos \hat \delta}{L \cosh y}\,.
\end{equation}
We can use these results to evaluate the $S_4$ surface  contribution given by
\begin{align}
\begin{split}\label{S4full}
 S_4 
= \frac{L}{8\pi\Gn} \int^{0}_{-y^*} dy \int_0^{\hat \delta} dt \cot \hat \delta
 = \frac{L}{8\pi\Gn}  y^*.
\end{split}
\end{align}

The additional lightcone surface $S_2$ generated from the boundary point at $\theta=0$ can be parameterized in terms of $t$ and $y$ as follows
\begin{align}\label{nullnull1}
x^{\mu}(t,y) = (t,y,\tanh^{-1} (\cos t)),
\end{align}
where $t \in [\hat \delta,\pi/2]$ and
$y \in [-y^*,0]$ and where $y=-y^*$ corresponds to a light ray which parallels the defect.
It is possible to verify, as we do below that this surface has zero null-expansion ($\Theta=0$) and as a result it is reparametrization invariant without the addition of the counterterm in eq.~\eqref{eq:CAaction} (in \cite{Lehner:2016vdi} this was referred to as a stationary hypersurface).\footnote{One way to understand this statement is that the surface $S_2$ is in fact a part of an entanglement wedge \cite{Czech:2012bh,Headrick:2014cta}. For the case of vacuum AdS and a spherical entangling surface, it is well known that the boundary of the entanglement wedge is a Killing horizon and the corresponding normals are null killing vectors \cite{Casini:2011kv,Faulkner:2013ica}. Hence this surface is known to have vanishing expansion and constant cross-sectional area when moving along its null generators.}
If we parameterize the surface with $\lambda$ such that $\lambda=\frac{L}{\mathcal{N}_2}\cosh(y)\ln\left(\tan \left(\frac{t}{2}\right)\right)$, we obtain for the normal vector\footnote{A guiding principle for this choice of parametrization is that it simplifies greatly the factor inside the logarithm, and as a consequence, the integration in the corner contributions $J_2$ and $J_3$.}
\begin{equation}\label{nullnull2}
k_{(2)}^{\mu} = \frac{dx^\mu}{d\lambda}= \frac{\mathcal{N}_2}{L \cosh y} \left(\sin t,0,-1\right),
\end{equation}
as well as the other properties of the surface $S_2$
\begin{equation}\label{nullnull3}
\gamma_{yy}= g_{\alpha\beta} e_y^\alpha e_y^\beta = L^2, \qquad
\Theta=0, \qquad
\kappa_{(2)} = - \frac{\mathcal{N}_2}{L} \frac{\cos t}{\cosh y}\, .
\end{equation}	
We can use these results to evaluate the surface contribution $S_2$ which reads
\begin{equation}\label{S2eqeq}
S_2	= \frac{L}{8\pi\Gn} \int_{-y^*}^{0}dy \int^{\pi/2}_{\hat{\delta}} dt \, \cot t
=-\frac{L}{8\pi\Gn}  y^* \ln\hat{\delta}\,,
\end{equation}
where in evaluating this expression we have changed the variable of integration from $\lambda$ to $t$ using the chain rule.
We proceed to evaluate the contribution of the joint $J_2$, associated with the surface of constant $t=\hat \delta$ and $y\in [-y^*,0]$
\begin{equation}\label{J2full}
J_2=-\frac{L\ln \mathcal{N}_2}{8\pi \Gn}  \int_{-y^*}^{0} \,\, dy =
-\frac{L\ln \mathcal{N}_2}{8\pi \Gn} y^*\,.
\end{equation}
The joint $J_3$, formed by the intersection of the two lightcone surfaces generated from the two antipodal points on the boundary, is characterized by $t=\pi/2$, $r = 0$ and $y\in[-y^*,0]$. At this intersection the normal vectors to the two null surfaces take the form
\begin{equation}
k_{(2)}^{\mu} = \frac{\mathcal{N}_2 }{L \cosh y} \left(1,0,-1\right),\qquad
\bar k_{(2)}^{\mu} = \frac{\mathcal{N}_2 }{L\cosh y} \left(1,0,1\right),
\end{equation}
which yields the following joint contribution
\begin{align}
J_3 =
\frac{L\ln \mathcal{N}_2}{4\pi \Gn} \int_{-y^*}^{0} \,\,dy =
\frac{L\ln \mathcal{N}_2}{4\pi \Gn} y^*\, .
\end{align}
Summing together all the contributions for the surfaces and joints inside the defect region we finally obtain
\begin{equation}
I_{\text{sj,in}}=8 (S_2+S_4+J_2)+4 J_3 =-\frac{L}{\pi \Gn} y^* \left(\ln\hat\delta-1\right),
\end{equation}
and of course, the parametrization freedom $\mathcal{N}_2$ canceled from this result.

\subsubsection{Defect Contribution}\label{subsec:defdef}
We now proceed to consider the defect contribution. We will include here both the brane action as well as the integration of the Einstein-Hilbert term over the infinitesimal thickness of the defect. The relation between these two contributions has been explored in \cite{Bachas:2002nz} using the Israel junction conditions \cite{Israel:1966rt}, where it was demonstrated that the Einstein-Hilbert contribution can be expressed in terms of the discontinuity of the extrinsic curvature across the defect, and this yields a contribution that is $(-2)$ times the brane action. Summing the two together results in a flipped sign for defect contribution
\begin{equation}\label{eq:defectcontri}
I_{d} = I_{\lambda} + I_{EH} = -I_{\lambda} = \lambda \int_{\text{defect}}  \sqrt{-h}=\frac{\tanh y^*}{4 \pi \Gn L} \int_{\text{defect}}  \sqrt{-h}
\end{equation}
where $h$ is the induced metric on the defect and we have used the relation \eqref{eq:defparam} to relate $\lambda$ and $y^*$. Of course, in the context of the CA conjecture we will be integrating over the part of the defect enclosed in the WDW patch. Since the defect lies inside the patch we do not need to add additional boundary contributions and joints at the location of the defect.\footnote{As an aside, we note that a naive extension of the prescription for joint terms between the defect surface and the additional $S_2$ boundary would fail, since in this case the null normal is included in the timelike defect surface which would result in a vanishing product of the normals to these two hypersurfaces.\label{strangeJoints}}
The defect brane corresponds to the constraint $y =-y^*$, see eq.~\eqref{eq:defb}.
We parameterize it by the coordinates $t$ and $r$, and its normal vector and induced metric are given by
\begin{equation}
\begin{split}
{\bf s^{(d)}} &\equiv s_{\mu}^{(d)}  dx^\mu = L \, dy, \\
{dh^2_{(d)}} &=
ds^2=  L^2 \cosh^2 y(-\cosh^2 r dt^2+ dr^2)\, .
\end{split}
\end{equation}
This yields the following defect contribution
\begin{equation}\label{CAdefects1234}
\begin{split}
 I_d &= \frac{L \sinh(2y^*)}{2\pi\Gn}
\int_{0}^{\tanh^{-1} (\cos \hat \delta)} dr \cosh r \int_{0}^{\cos^{-1}(\tanh r)} dt =  -\frac{L\sinh(2 y^*)}{2\pi\Gn}  \left( \ln \hat{\delta} -1\right),
\end{split}
\end{equation}
where we have included an overall factor of two to account for the future and past portions of the defect brane.

\subsubsection{Total CA Contribution}
We can now collect all the terms to obtain the total result for CA complexity using eq.~\eqref{eq:CAformula}\footnote{We have set $\hbar=1$.}
\begin{equation}\label{final-result-CA}
\begin{split}
\mC_A &={  I_{\mt{WDW}} \over \pi} ={  1 \over \pi} \left( I_d+I_{\text{sj,in}}+I_{\text{sj,out}}+I_{\text{bulk}}\right)
=   \frac{c_{{}_T}}{  3\pi } \left(\frac{1}{\hat \delta}\left[\ln\left(\frac{\ell_{ct}}{L}\right)+1\right] +  \frac{\pi}{2}\right),
\end{split}
\end{equation}
where we have expressed the result in terms of the central charge $c_T = 3 L/ (2 \Gn)$ of the boundary theory.
We see that the presence of the defect does not change the result! This is in contrast to the logarithmic contribution introduced into the CV complexity due to the presence of the defect, cf.~eq.~\eqref{eq:CVresult}. There is an ambiguity related to the new scale $\ell_{ct}$ introduced by the counterterm which has been suggested to be related to certain choices that can be made in defining the complexity in the QFT side \cite{Jefferson:2017sdb,Chapman:2017rqy}, see section 5 of \cite{Chapman:2018lsv}.

\section{Holographic Complexity for Subregions}\label{sec:subregion}
Next, we investigate extensions of the CV and CA conjectures for mixed states produced by tracing out the degrees of freedom outside a subregion $A$ of the full boundary time slice, see \cite{Alishahiha:2015rta,Carmi:2016wjl}. Both proposals are motivated by the suggestion that the natural bulk region encoding the reduced density matrix is the entanglement wedge  \cite{Czech:2012bh,Headrick:2014cta}. In the presence of the defect, the non-trivial case is when the subregion $A$ includes   the defect and we focus on this case below.

\subsection{Subregion CV Conjecture}
\begin{figure}
	\centering
	\includegraphics[scale=0.4]{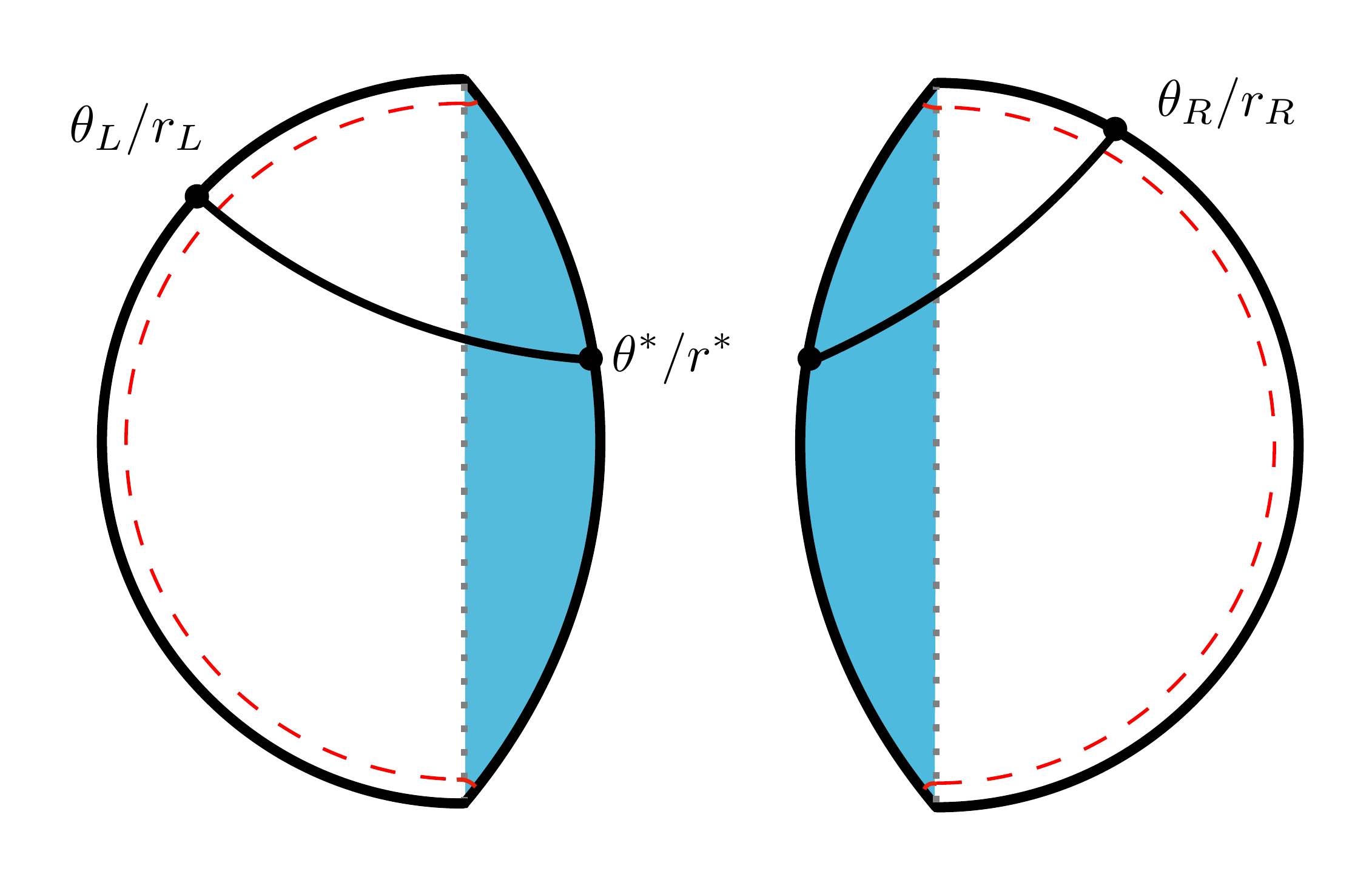}
	\caption{Defect AdS geometry consisting of two AdS patches glued together along lines of constant $y=\pm y^*$ at the location of the defect. The spacelike geodesic connecting $\theta_L$ and $\theta_R$ (alternatively $r_L$ and $r_R$) will pass through $\theta=\pm\theta^*$ (alternatively $r=r_*$) on the left/right patches respectively. On the right patch we have  $-y^*\leq y <\infty$ while on the left patch we have $-\infty < y \leq y^*$. We have extended the definition of $r$ in both patches such that all the patch is covered and $-\infty \leq r \leq \infty$. Angles are measured with respect to the vertical upward direction.}
	\label{fig:sCV_whole}
\end{figure}

The extension of the CV conjecture for the complexity of mixed states \cite{Alishahiha:2015rta,Carmi:2016wjl} suggests that the complexity is proportional to the maximal volume of a codimension-one surface enclosed between the boundary region $A$ and its corresponding Ryu-Takayanagi (RT) surface \cite{Ryu:2006bv,Ryu:2006ef} with the same proportionality coefficient as in equation \eqref{eq:CVformula}. We will use this prescription in the defect-AdS geometry for the case in which the subregion $A$ contains the defect.
For this purpose, we need to find the RT surface (the spacelike geodesic, in our case) connecting two points on opposite sides of the defect as illustrated in figure \ref{fig:sCV_whole}. This is done by matching two geodesics connecting the two boundary points on each side of the defect to the same point on the defect surface and minimizing the total length.

\subsubsection{Finding the Geodesics}
Let us start with the metric on a constant time slice in global coordinates on the right patch, see eq.~\eqref{eq:metric.t.phi.theta}
\begin{equation}
ds^2  = {L^2 \over \cos^2\phi} \left( d\phi^2 +\sin^2 \phi \, d\theta^2\right).
\end{equation}
The geodesics for this metric can be found by minimizing the line element. This leads to the following geodesic equation parameterized by $\theta$, where we have used the change of variables $\Phi(\theta) = \sin (\phi(\theta))$
\begin{equation}
-\Phi(\theta) \Phi''(\theta)+2 \Phi'(\theta)^2+\Phi(\theta)^2=0\, ,
\end{equation}
which admits the general solution
\begin{equation}\label{geo-solsol22}
  \sin \phi\, \cos(\theta-\alpha) = c \,.
\end{equation}
$\alpha$ and $c$ are two constants of integration which will be fixed by the boundary conditions $\theta_L$ and $\theta_R$ where the geodesic meets the boundary of AdS. This demonstrates that these are simply curves of constant $r$, rotated by an angle $\alpha$ cf.~eq.~\eqref{eq:coorconvert2}.

Alternatively, we can work with the $y$ and $r$ coordinates by extending the definition of $r$ to negative values, in order to cover the full space. This is done by formally extending the coordinate transformation in eq.~\eqref{eq:coorconvert2} to angles $\theta>\pi/2$ or $\theta<-\pi/2$. This choice of coordinates turns out to be the most convenient when evaluating the relation between the integration constants and the boundary conditions of the geodesics.
The geodesic in the $y$ and $r$ coordinates can be obtained by considering the restriction of the metric \eqref{DCFTsol2} onto a constant time slice
\begin{equation}\label{Pizza2}
ds^2  = L^2 \left( \left(\frac{dy}{dr}\right)^2 +\cosh^2 y\right) dr^2 .
\end{equation}
This leads to the following geodesic equation where we have used the change of variables $Y(r)=\tanh (y(r))$
\begin{equation}
\frac{d^2 Y}{d r^2}-Y=0.
\end{equation}
This equation admits the general solution
\begin{equation}\label{geo-solsol22}
 \tanh(y(r)) =  c_1 e^r+c_2 e^{-r}
\end{equation}
where $c_1$ and $c_2$ will be fixed by the boundary conditions. In general these constants will be different for the left and right sections of the geodesic and we will have to match them at the position where the geodesic meets the defect. Fixing the boundary conditions $y=\infty$, $r=r_R$ for the right section of the geodesic and $y=-\infty$, $r=r_L$ for the left section we can express the geodesic solutions as follows
\begin{equation}
\begin{split}\label{eq:geoyrd}
\sinh(r-r_R +\tanh^{-1} (a_R)) &= \frac{a_R}{\sqrt{1-a_R^2}}\tanh y,\\
\sinh(r-r_L -\tanh^{-1} (a_L)) &= \frac{a_L}{\sqrt{1-a_L^2}}\tanh  y,
\end{split}
\end{equation}
where the constants of integration $a_L$ and $a_R$ will be fixed by matching the two geodesics on the two sides of the defect.

Since the metric is continuous at the location of the defect (only its derivative with respect to the $y$ coordinate is discontinuous), one can show by integrating the equations of motion in a small pillbox around the defect that $dy/dr$ is continuous at the point where the geodesics cross the defect. This is a local matching condition which is equivalent  to minimizing the total length of the geodesics. Explicitly, the matching condition reads
\begin{equation}
\frac{\sqrt{1-a_R^2}}{a_R} \cosh(r_*-r_R +\tanh^{-1} (a_R))=\frac{\sqrt{1-a_L^2}}{a_L} \cosh(r_*-r_L -\tanh^{-1} (a_L))
\end{equation}
where $r=r_*$ is the value of $r$ at the point where the geodesics cross the defect. In addition, the fact that the geodesics in eq.~\eqref{eq:geoyrd} cross the defect at $r=r_*$ yields the following conditions
\begin{align}
\sinh(r_*-r_R +\tanh^{-1} (a_R)) &= -\frac{a_R}{\sqrt{1-a_R^2}}\tanh y^*,\\
\sinh(r_*-r_L -\tanh^{-1} (a_L)) &= \frac{a_L}{\sqrt{1-a_L^2}}\tanh y^*.
\end{align}
Solving these three equations leads to
\begin{equation}\label{eqfora}
a \equiv a_L=a_R = \frac{\sinh(\frac{r_R-r_L}{2})}{\cosh(\frac{r_R-r_L}{2})+\tanh y^* }, \qquad r_* = \frac{r_L+r_R}{2}.
\end{equation}
We note that the point $r_*$ is simply the arithmetic mean of the two asymptotic values of $r$ on the two sides of the defect. We also note that $|a|<1$, and the sign depends on whether $r_R>r_L$ or $r_L>r_R$.

\subsubsection{Evaluating the Volume}
\begin{figure}
\centering
\includegraphics[scale=0.3]{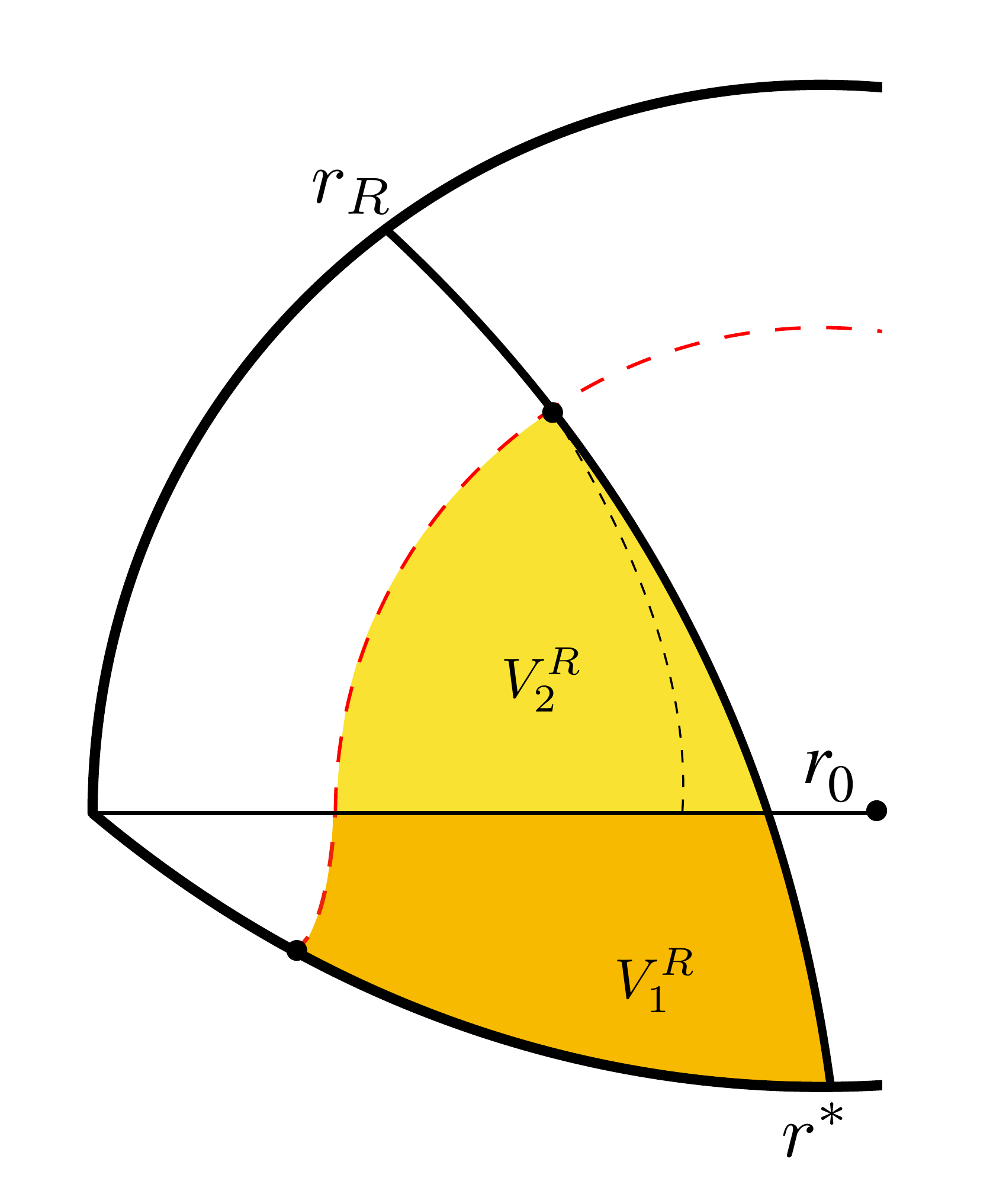}
\caption{A corner of the right defect patch illustrating the relevant volumes in the evaluation of the subregion CV proposal. The red dashed curve indicates the cutoff surface and the volumes $V_1^R$ in eq.~\eqref{eq:v1R} and $V_2^R$ in eq.~\eqref{eq:v2R} are colored in dark and light yellow respectively. The dashed black line indicates the division between the two integration regions in eq.~\eqref{eq:v2R}.}
\label{fig:sCV-corner}
\end{figure}

We are now in the position to evaluate the volume enclosed inside the geodesic studied in the previous subsection as suggested by the CV proposal. We have divided the volume to the part inside the defect region and the part outside the defect region. Throughout the calculation we have assumed that $r_L,r_R\ll\ln(2/\hat\delta)$, namely that the size of the boundary interval as well as the distance between its end points and the defect are kept finite and far below the cutoff value. The volume of the part inside the defect region on the right patch can be evaluated as
\begin{equation}\label{eq:v1R}
V_1^R=   L^2  \int_{-y^*}^{0} dy \cosh y
\int_{r_R -\tanh^{-1}(a)+\sinh^{-1}\left(\frac{a}{\sqrt{1-a^2}}\tanh y \right)}^{\tanh^{-1}(\cos \hat \delta)} dr
\end{equation}
and for the left patch we have
\begin{equation}
V_1^L=  L^2  \int_{0}^{y^*} dy \cosh y \int_{r_L+\tanh^{-1}(a)+\sinh^{-1}\left(\frac{a}{\sqrt{1-a^2}}\tanh y \right)}^{\tanh^{-1}(\cos \hat \delta)} dr.
\end{equation}
Using the change of variables $y\rightarrow-y$ in the first integral, we can combine the two integrals.
Some of the contributions cancel out and we are left with
\begin{equation}
\begin{split}
V_1 = & \, V_1^L+V_1^R=   L^2  \int_{0}^{y^*} dy \cosh y \left(2\tanh^{-1}(\cos \hat \delta)-r_L-r_R\right)
\\
= & \, \sinh y^*  \left(2\ln \left(\frac{2}{\hat \delta}\right)-r_L-r_R\right).
\end{split}
\end{equation}
The volume outside the defect region for the right patch is given by
\begin{align}\label{eq:v2R}
V_2^R&= L^2 \int_{r_R+\mathcal{O}(\hat \delta^2)}^{\tanh^{-1}(\cos \hat\delta )} dr
\int_{0}^{\cosh^{-1}\left(\frac{1}{\cosh r \sin \hat{\delta} }\right)} \cosh  y \,  dy\no\\
&~~~~ + L^2 \int^{r_R+\mathcal{O}(\hat \delta^2)}_{r_0} dr \int_{0}^{\tanh^{-1}\left(\frac{\sqrt{1-a^2}}{a}\sinh\left(r-r_R +\tanh^{-1} (a)\right)\right)} \cosh  y \, dy \no\\
&=L^2\left( \frac{1}{\hat{\delta}}\cos^{-1} (\tanh r_R)-\pi/2\right)  + L^2 \arcsin(a),
\end{align}
where we have decomposed the volume integration into two parts along the black dashed line in figure \ref{fig:sCV-corner}, and the first integral was carried out using the change of variables $t=\sinh r $. We have also defined $r_0= r_R - \tanh^{-1}(a)$ which is the point on the spatial geodesic \eqref{eq:geoyrd} where $y=0$. Note that in some cases  the limits of integration  in the second integral may be flipped which accounts for a subtraction rather than an addition.
The volume in the left patch can be effectively obtained by replacing $a\to-a$ and $r_R\to r_L$ in the above expression which yields
\begin{equation}
V_2 = V_2^R+V_2^L = \frac{L^2}{\hat{\delta}}\left( \cos^{-1} (\tanh r_R)+\cos^{-1} (\tanh r_L)\right)   -L^2 \pi.
\end{equation}

Finally summing the different contributions yields the following result for the subregion complexity using the CV conjecture
\begin{equation}\label{eq:subCVfinal}
\mC^{\text{sub}}_V(r_R,r_L)= {2 c_T \over 3} \left(\frac{\theta_R-\theta_L}{\hat\delta} + \sinh y^*  \left(2\ln \left( {2 \over\hat\delta} \right)- r_L- r_R  \right) - \pi \right),
\end{equation}
where we have expressed the result in terms of the central charge $c_T=3L/(2\Gn)$ and the opening angle $\theta_R-\theta_L$ where
\begin{equation}\label{thetaLthetaR}
\theta_{R}=\cos^{-1}(\tanh  r_{R} ),\qquad \theta_{L}=-\cos^{-1}(\tanh  r_{L} ),
\end{equation}
cf.~\eqref{eq:coorconvert2} with $\phi=\pi/2$.
One consistency check on our result is to check that when $r_L=r_R=0$ we recover half the volume of the full time slice, which is indeed the case, cf. eq.~\eqref{eq:CVresult}.
The leading divergence in eq.~\eqref{eq:subCVfinal} is proportional to the size the interval $A$ measured in terms of its opening angle $\theta_R-\theta_L$, which is the same result as obtained without the defect, see \cite{Carmi:2016wjl}. The last term $-\pi$ is a topological term, already mentioned in reference \cite{Abt:2017pmf}. There, the authors concluded that the holographic subregion complexity of $q$ intervals living on the boundary of AdS$_3$ is proportional to $\frac{x}{\delta}+\pi q - 2 \pi \chi$,
where $x$ is the total length of the entangling intervals on the boundary and $\chi$ is the Euler characteristic of the codimension-one volume entering in the CV proposal. In our case, we obtain exactly the same result for the theory without a defect by setting $y^*=0$, with $\chi=1$ and $q=1$ (or alternatively $q=0$ for the full boundary, cf.~eq.~\eqref{eq:CVresult}).
Compared to the full CV calculation in eq.~\eqref{eq:CVresult}, the subregion complexity has half of the log divergent contribution which is due to the fact that the subregion encloses only one boundary defect and the finite piece has an additional negative contribution proportional to $r_L+r_R$.

Finally, with the tools we have developed here we can also generalize the result of \cite{Azeyanagi:2007qj} for the entanglement entropy in the presence of the defect to the case of an entangling region which is not symmetric around the defect. The entanglement entropy is determined by the minimal area surface anchored at the boundary of the entangling region according to the Ryu-Takayanagi (RT) formula  $S_{EE} = A/ (4 \Gn)$ \cite{Ryu:2006bv,Ryu:2006ef}, where for AdS$_3$, $A$ is simply the length of the geodesic \eqref{eq:geoyrd} according to the length element in eq.~\eqref{Pizza2}. In total we have
\begin{equation}
S_{EE}=S_{EE,\text{empty}}+\Delta S_{EE,\text{defect}}
\end{equation}
where the entropy in the absence of the defect is given by \cite{Calabrese:2004eu}
\begin{equation}
S_{EE,\text{empty}}= \frac{c_T}{3} \ln \left(\frac{2 \sin\left(\frac{\theta_R-\theta_L}{2}\right)}{\hat \delta}\right)
\end{equation}
and the entropy associated with the defect is given by
\begin{equation}\label{EEdefectsub}
\Delta S_{EE,\text{defect}} =\frac{c_T}{3} \ln\left(  \cosh y^*+ \frac{\sinh y^*}{\cosh\left({r_R-r_L \over 2}\right)}  \right)  \,.
\end{equation}
For the case $r_R=r_L=0$ where the geodesic passes through the center of the AdS$_3$ this matches eq.~(3.9) of \cite{Azeyanagi:2007qj}.
In fact, the authors there note that as long as the entangling surface is symmetric around the defect, the result does not depend on the size of the subsystem, and is related by means of a folding trick to the boundary entropy $\ln g$. Indeed we observe that when setting $r_L=r_R$ the dependence on the boundary points $r_L$, $r_R$ disappears from the above equation. If the defect is not located at the midpoint of the interval, the entanglement entropy is no longer determined solely by the two universal numbers $c_T$ and $g$ but rather depends also on the location of the end points of the entangling region.

\subsection{Subregion CA Conjecture}\label{subcasecsec}
In \cite{Carmi:2016wjl}, a proposal was made for extending the CA conjecture to subregions (corresponding to mixed states); the proposal is that the complexity of the mixed state is proportional to the action of a codimension-zero bulk region, defined as the intersection of the WDW patch and the entanglement wedge associated to the relevant subregion with the proportionality coefficient as in eq.~\eqref{eq:CAformula}. The WDW patch does not depend on the subregion and is therefore identical to the one described in subsection \ref{sec:WDWpatch}. The entanglement wedge associated to a boundary subregion $A$ is the set of bulk points which are spacelike separated from the RT surface and connected to the boundary domain of dependence of the subregion $A$, see \cite{Czech:2012bh,Headrick:2014cta}. Its boundary is then formed by the light-front of the past and future light cones emanating from the various points on the RT surface. For the case of vacuum AdS$_3$ the null geodesics which form the boundaries of the entanglement wedge meet on the boundary at the two ends of the causal diamond associated with the subregion $A$.

\begin{figure}
\begin{center}
\includegraphics[width=0.35\textwidth]{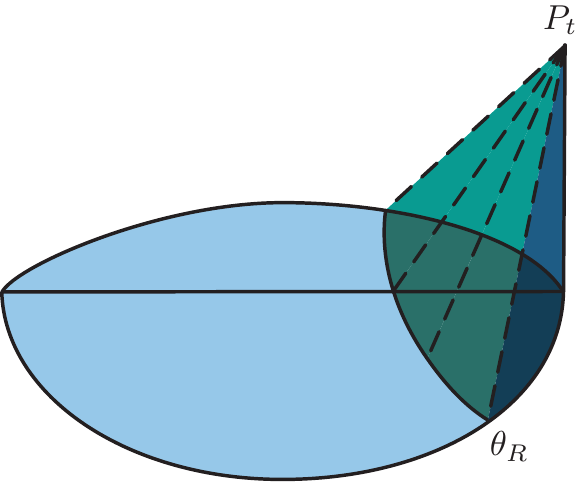}
\end{center}
\caption{Illustration of the entanglement wedge for a subregion centered around the defect in the right patch of the defect AdS$_3$ geometry. $P_t$ is the point on the boundary at the edge of the causal diamond associated with the relevant boundary region, whose past bulk lightcone will pass through the spatial geodesic connecting $\theta_L$ and $\theta_R$.}
\label{fig:Ew2}
\end{figure}

To simplify the calculation we will be focusing on the case where the entangling region is symmetric about the defect, i.e., $r_R=r_L$ or $\theta_L+\theta_R=0$, see eq.~\eqref{thetaLthetaR}. In this case, the RT surface is simply a curve of constant $r=r_L=r_R$, see eqs.~\eqref{eq:geoyrd} and \eqref{eqfora}. The entanglement  wedge then naturally coincides with the one of empty AdS$_3$ and consists of the light rays emanating from the boundary point $P_t = (t,\phi,\theta) = (\theta_R,\pi/2,0)$, see figure \ref{fig:Ew2}. In fact, recall that we have already considered a similar lightcone, when we were looking at the extension of the boundary of the WDW patch in the defect region in subsection \ref{sec:WDWpatch}, where it was described by the relation \eqref{lightconeryt}. Adapting this expression to our case by the substitution $t\rightarrow \theta_R-t$ results in the following parametrization for the boundary of the entanglement wedge
\begin{equation}\label{eq:BEWsymm}
\tanh r  = \cos(\theta_R-t)\, .
\end{equation}
For the case of vacuum AdS and spherical entangling regions, it is well known that the boundary of the entanglement wedge is a Killing horizon which has vanishing expansion \cite{Casini:2011kv,Faulkner:2013ica} and therefore  the counterterm in eq.~\eqref{eq:CAaction} will vanish for this surface.

In the following, we will divide the contributions to the CA proposal for the subregion to two parts --- inside, and outside the defect region
\begin{equation}
\mC_{A,\text{sub}} = \mC_{A,\text{sub}}^{\text{vac}} + \mC_{A,\text{sub}}^{d}.
\end{equation}
Since we are mainly interested in studying the special properties that the defect induces in our system, we will focus here on evaluating the contributions to the complexity from the defect region. Those will be the ones important for the conclusions of this paper. For completeness we also extract the divergent contributions outside the defect region in appendix \ref{app:sCAallpieces}. We will demonstrate below that $\mC_{A,\text{sub}}^{d}$ vanishes for all symmetric subregions around the defect.

In what follows it will be useful to have an explicit expression for the intersection of the WDW patch and the entanglement wedge in the defect region. Combining \eqref{eq:BEWsymm} and \eqref{lightconeryt} for this joint yields
\begin{equation}\label{ttRhalf}
t=\theta_R/2\, .
\end{equation}

\subsubsection{Evaluating the Action}
\begin{figure}
	\centering
	\begin{subfigure}[b]{0.31\textwidth} 
		\centering \includegraphics[width=\textwidth]{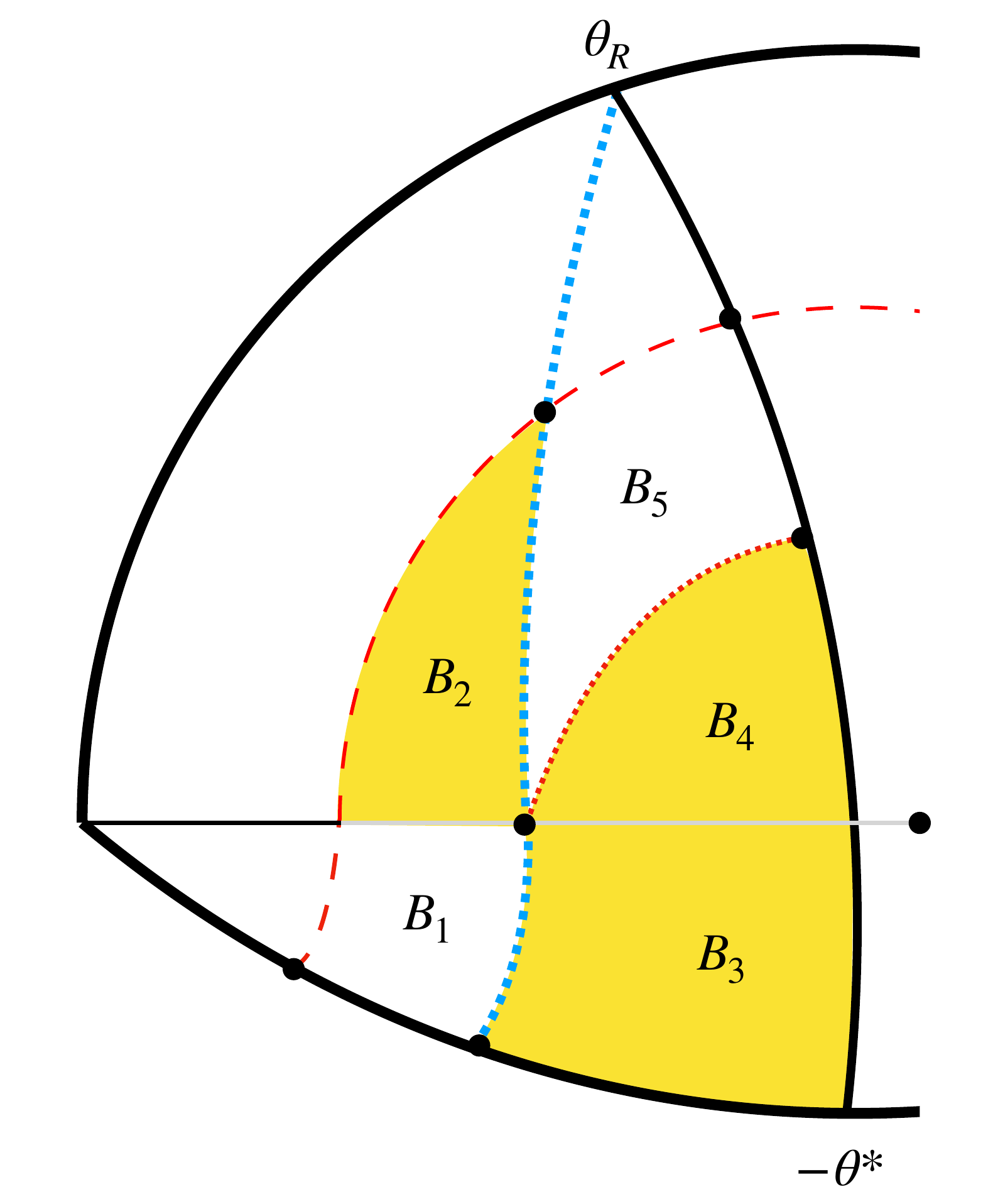}
		\caption{Bulk contributions}\label{fig:sCA_vol}
	\end{subfigure}
	~~~~ 
	\begin{subfigure}[b]{0.29\textwidth}
		\centering \includegraphics[width=\textwidth]{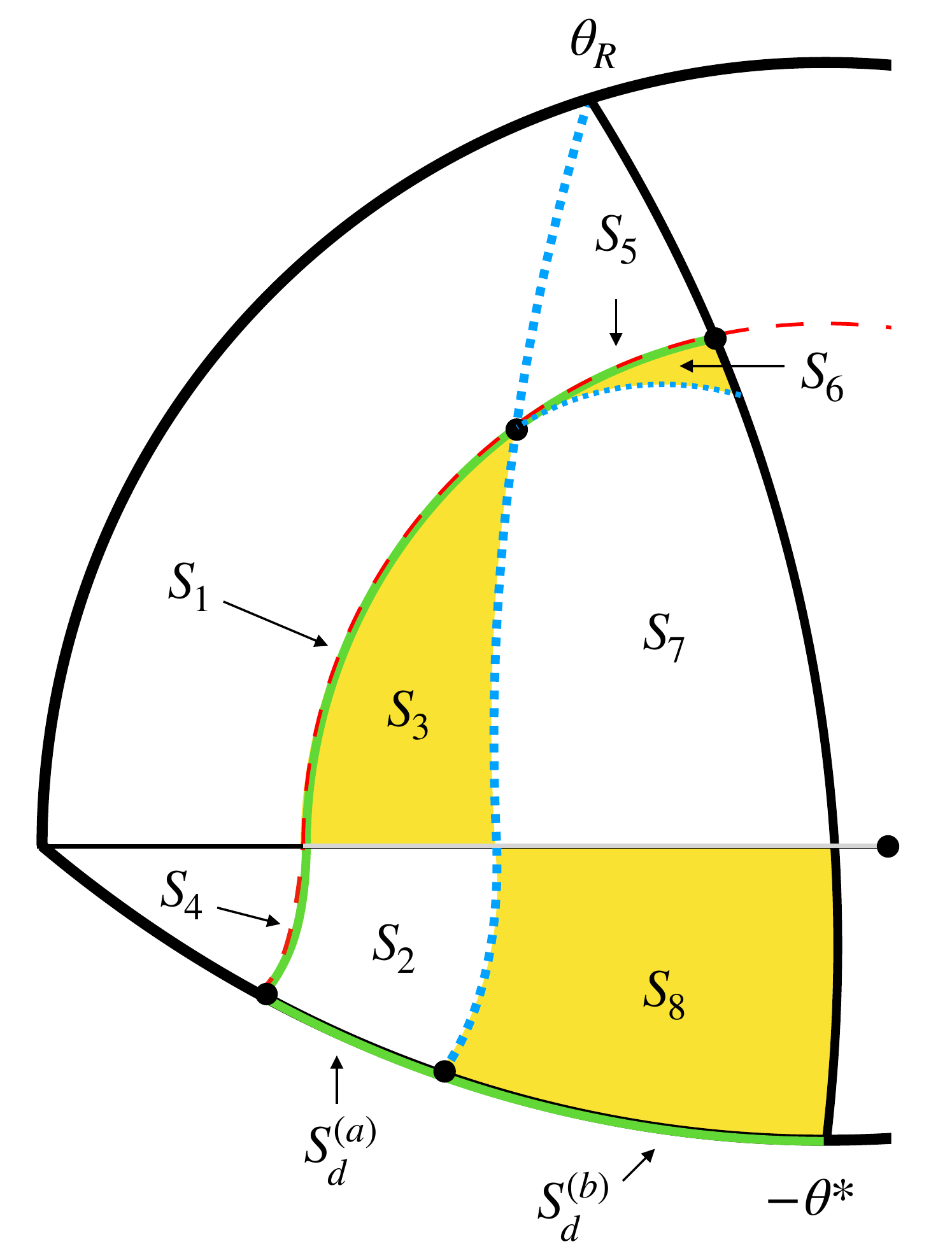}
		\caption{Surface contributions}\label{fig:sCA_surf}
	\end{subfigure}
	~~~
	\begin{subfigure}[b]{0.29\textwidth}
		\centering \includegraphics[width=\textwidth]{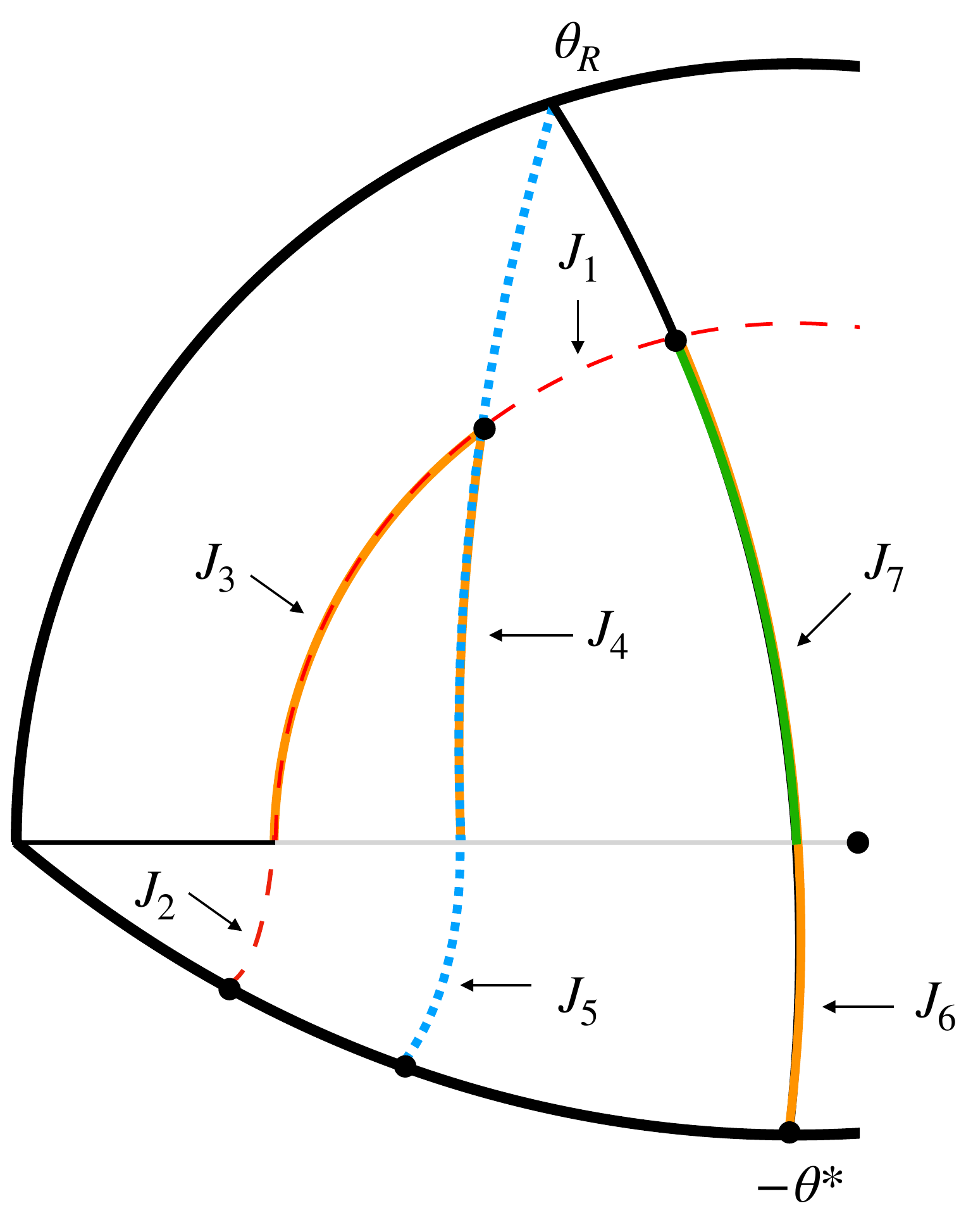}
		\caption{Joint contributions}\label{fig:sCA_joint}
	\end{subfigure}
	\caption{Illustrations of the various contributions in the evaluation of the subregion CA proposal for an interval which is symmetric around the defect. The illustrations focus on the right patch, but of course, equivalent contributions exist for the left patch. The red dashed line represents the cutoff and the middle dotted blue curve represents the projection of the joint formed at the intersection between the boundary of the WDW patch and the entanglement wedge. We have also included certain internal divisions between $B_4$ and $B_5$ and between  $S_6$ and $S_7$ outside the defect region which we use in evaluating the relevant integrals in appendix \ref{app:sCAallpieces}.}
	\label{fig:sCA}
\end{figure}

In this subsection we focus on contributions from the defect region. We quote the result for the structure of divergences outside the defect region at the end of the subsection and the details can be found in appendix \ref{app:sCAallpieces}.
The projections of the various relevant contributions onto the $t=0$ time slice are illustrated in figure \ref{fig:sCA}. They consist of bulk, boundary, joint and defect contributions. In the defect region, those are the two bulk contributions $B_1$ (region under the WDW patch) and $B_3$ (region under the entanglement wedge), the three surface contributions $S_4$ (cutoff surface), $S_2$ (null boundary of the WDW patch) and $S_8$ (null boundary of the entanglement wedge), the three joint contributions $J_2$ (between the cutoff surface and the boundary of the WDW patch), $J_5$ (between the boundary of the WDW patch and the boundary of the entanglement wedge) and $J_6$ (between the past and future boundaries of the entanglement wedge), and the two defect contributions $S_d^{(a)}$ (enclosed under the WDW patch) and $S_d^{(b)}$ (enclosed under the entanglement wedge). We evaluate them below.

\paragraph{Bulk contributions}
We start from the bulk contribution $B_1$, bounded by the WDW patch  \eqref{lightconeryt}, which reads
\begin{align}
	B_1&= -{L \over 4\pi \Gn } \int_{-y^*}^{0}\ch^2 y \, dy\int_{\tanh^{-1}(\cos{\theta_R \over 2})}^{\tanh^{-1}(\cos\hat{\delta})} \ch r\, dr \int_{0}^{\cos^{-1}(\tanh r)}dt\no\\
	&=  {L \over 8\pi \Gn }\left(y^*+\frac{1}{2}\sinh(2 y^*)\right)
\left(\ln \hat\delta +\frac{\theta_R}{2}  \cot \left(\frac{\theta_R}{2}\right)-\ln \left(\sin 	 \left(\frac{\theta_R}{2}\right)\right)-1\right).
\end{align}
Next, we evaluate the bulk contribution $B_3$, under the entanglement wedge \eqref{eq:BEWsymm}, which reads
\begin{align}
B_3 & =  -{L \over 4\pi \Gn } \int_{-y^*}^{0}\ch^2 y \, dy\int^{\tanh^{-1}(\cos{\theta_R \over 2})}_{\tanh^{-1}(\cos{\theta_R })} \ch r\, dr \int_{0}^{\theta_R - \cos^{-1}(\tanh r)}dt\no\\
&= -{L \over 8\pi \Gn }\left(y^*+\frac{1}{2}\sinh(2 y^*)	\right) \left( \frac{\theta_R}{2} \cot \left(\frac{\theta_R}{2}\right)+\ln \left(\sin \left(\frac{\theta_R}{2}\right) \csc \theta_R \right)\right).
\end{align}
Multiplying by four to account for the equivalent contributions from both sides of the defect, as well as above and below the $t=0$ time slice, we find that the total bulk contribution inside the defect region is given by
\begin{align}\label{eq:sCAB}
	I_{\text{bulk,in}}=  {L \over 2\pi \Gn }
\left(y^*+\frac{1}{2}\sinh(2 y^*)\right)
\left(\ln \hat\delta -\ln \left(\frac{1}{2}\tan \left(\frac{\theta_R}{2}\right)\right)-1\right)\,.
\end{align}

\paragraph{Surface and joint contributions}
The contribution from the cutoff surface $S_4$ has already been evaluated in subsection \ref{sec:CAsurfaces}, see eq.~\eqref{S4full}, and is given by
\begin{equation}
	S_4= {L \over 8\pi \Gn }y^*.
\end{equation}
The contribution of the null surface $S_2$ is very closely related to the one evaluated in eq.~\eqref{S2eqeq}. All one has to do is modify the limits of integration according to eq.~\eqref{ttRhalf} which yields
\begin{equation}
S_2	= \frac{L}{8\pi\Gn} \int_{-y^*}^{0}dy \int^{\theta_R/2}_{\hat{\delta}} dt \, \cot t
=\frac{L}{8\pi\Gn}  y^* \left(-\ln \hat{\delta}+\ln\left(\sin \left(\frac{\theta_R}{2}\right)\right)\right)\,.
\end{equation}
The details of the null boundary of the entanglement wedge $S_8$ can be easily obtained from those of the null boundary of the WDW patch in eq.~\eqref{nullnull1}-\eqref{nullnull3} by substituting $t\rightarrow   \theta_R-t$ in the relevant places. However, we have to make sure that the parametrization $\lambda$ increases from past to future, hence we choose  $\lambda=-\frac{L}{\mathcal{N}_\mt{EW}}\cosh y\,\ln\left(\tan \left(\frac{\theta_R-t}{2}\right)\right)$, where we have included a constant $\mathcal{N}_\mt{EW}$ to account for the choice of parametrization at the boundaries of the entanglement wedge. The surface data is given by
\begin{equation}\label{dataEW}
\begin{split}
&  x^{\mu}(t,y) = (t,y,\tanh^{-1} (\cos (\theta_R-t))),
\quad
k_{(2)}^{\mu} = \frac{\mathcal{N}_\mt{EW}}{L \cosh y } \left(\sin(\theta_R-t),0,1\right), \\
& \gamma_{yy} = L^2, \qquad
\Theta=0, \qquad
\kappa_{(2)} =  \frac{\mathcal{N}_\mt{EW}}{L} \frac{\cos(\theta_R-t)}{\cosh y },
\end{split}
\end{equation}
which leads to the following surface contribution
\begin{equation}\label{S8subCA}
\begin{split}
S_8	= &\,-\frac{L}{8\pi\Gn} \int_{-y^*}^{0}dy \int^{\theta_R/2}_{0} d t \, \cot (\theta_R-t)
=-\frac{L}{8\pi\Gn} y^* \ln \left(2 \cos \left(\frac{\theta_R}{2}\right)\right)\,.
\end{split}
\end{equation}

Next, we evaluate the relevant joints. The joint $J_2$ at the intersection of the WDW patch and the cutoff surface is identical to the one evaluated in eq.~\eqref{J2full} and reads
\begin{equation}
J_2=
-\frac{L\ln \mathcal{N}_2}{8\pi \Gn} y^*\,.
\end{equation}
The joint $J_5$ at the intersection of the WDW patch and the entangling wedge can be obtained using the normal vectors in eqs.~\eqref{dataEW} and \eqref{nullnull2} evaluated at $t=\theta_R/2$ which yields
\begin{equation}
J_5=\frac{L\ln \left(\mathcal{N}_\mt{EW}\mathcal{N}_2\right)}{8\pi \Gn} y^*\, .
\end{equation}
The joint $J_6$ is obtained by contracting the normal vector in eq.~\eqref{dataEW} for $t=0$ with the normal vector of the past null boundary of the entanglement wedge obtained from the former by flipping the sign of its $t$ component. This yields
\begin{equation}\label{J6d}
J_6=-\frac{L\ln \mathcal{N}_\mt{EW} }{4\pi \Gn} y^*\, .
\end{equation}
Summing all these contributions together yields the following result for the action of the surfaces and joints  inside the defect region
\begin{equation}\label{subsjin}
\begin{split}
I_{\text{sj,in}}=&\,4(S_2+S_4+S_8+J_2+J_5)+2J_6
\\
=&\,
\frac{L}{2\pi\Gn} y^* \left(-\ln \hat{\delta} +\ln \left( \frac{1}{2}\tan \left(\frac{\theta_R}{2}\right)\right)+1\right)
\end{split}
\end{equation}
and of course, we note that the parametrization choices $\mathcal{N}_2$ and $\mathcal{N}_\mt{EW}$ canceled out.

\paragraph{Defect contribution}
The defect contribution is given according to eq.~\eqref{eq:defectcontri}. One has to subdivide the integration into two parts. First, we consider the defect brane portion $S_d^{(a)}$ under the WDW patch
\begin{equation}
\begin{split}
S_d^{(a)} &= {L \sh  2y^*\over 8\pi \Gn }\int_{\tanh^{-1}(\cos{\theta_R \over 2})}^{\tanh^{-1}(\cos\hat{\delta})} \ch r dr \int_{0}^{\cos^{-1}(\tanh r)}dt\no\\
&= {L \sh  2y^*\over 8\pi \Gn }\left(-\ln \hat\delta -\frac{\theta_R}{2}  \cot \left(\frac{\theta_R}{2}\right)+\ln \left(\sin
\left(\frac{\theta_R}{2}\right)\right)+1\right).
\end{split}
\end{equation}
Next we evaluate the defect brane contribution under the entanglement wedge
\begin{equation}
\begin{split}
S_d^{(b)} &={L \sh  2y^*\over 8\pi \Gn }\int^{\tanh^{-1}(\cos{\theta_R \over 2})}_{\tanh^{-1}(\cos{\theta_R })} \ch r dr \int_{0}^{\theta_R - \cos^{-1}(\tanh r)}dt\no\\
&={L \sh  2y^*\over 8\pi \Gn } \left( \frac{\theta_R}{2}  \cot \left(\frac{\theta_R}{2}\right)+\ln \left(\sin \left(\frac{\theta_R}{2}\right) \csc \theta_R\right)\right).
\end{split}
\end{equation}
Those contributions are counted twice to account for the parts of the defect brane to the future and past of the $t=0$ time slice. Finally, we obtain
\begin{align}\label{eq:sCAD}
	I_d &=2\left(S_d^{(a)}+S_d^{(b)}\right)={L  \sinh(2y^*)\over 4\pi \Gn }\left(-\ln  \hat\delta +\ln \left(\frac{1}{2}\tan \left(\frac{\theta_R}{2}\right)\right)+1\right).
\end{align}

\paragraph{Total contributions from defect region}
Adding up the contributions \eqref{eq:sCAB}, \eqref{subsjin} and \eqref{eq:sCAD}, we find that the defect region contribution to the subregion complexity vanishes
\begin{equation}\label{eq:subCAdresult}
	\mC_{A,\text{sub}}^{d} =\frac{1}{\pi}\left(I_{\text{bulk,in}}+I_{\text{sj,in}}+I_d\right)=0
\end{equation}
as was the case for the complexity of the state on the entire time slice. This means that the subregion complexity for an interval centered around the defect  will be identical to the result for a subregion of the same size in empty AdS. This is again in stark contrast to the results of the subregion CV complexity in eq.~\eqref{eq:subCVfinal}, where the defect introduced a logarithmic contribution which also depended on the location of the end points of the subregion.

\paragraph{Contributions outside the defect region}
In appendix \ref{app:sCAallpieces} we consider the contribution to the complexity from outside the defect region. This is the same as evaluating the subregion complexity for empty AdS.\footnote{For the case of a flat boundary, the divergence structure of the subregion complexity in vacuum AdS was studied in \cite{Carmi:2016wjl}.}
We are able to extract the structure of divergences analytically and obtain
\begin{equation}\label{toderive}
\mathcal{C}_{A,\text{sub}}^\text{vac}  =
{c_T\over  3\pi^2}\left({\theta_R \over \hat{\delta}}\left[ \ln \left( {\ell_{ct}  \over L} \right) + 1 \right] + \ln \hat{\delta} \,\ln\left(\frac{2\ell_{ct}}{L}\right) \right)+\text{finite}.
\end{equation}
We note that upon setting $\theta_R=\pi$ we recover the leading divergence of the full boundary complexity \eqref{final-result-CA}. However,
note that in expanding this result we have everywhere assumed that $\theta_R$ was not too close to the cutoff, and therefore we cannot expect to recover the subleading divergences in the full boundary complexity in this way. We see that the result here has an additional logarithmic contribution compared to that in eq.~\eqref{final-result-CA} which depends on the scale $\ell_{ct}$ associated with the counter term.

\section{Complexity in QFT}\label{sec:QFT}
In this section we consider the problem of calculating the defect contribution to the complexity of the ground state from the dual field theory point of view.

At the moment it is not known, even in principle, how one should compute the complexity for a generic interacting field theory, although some progress has been made for weakly interacting QFTs, see \cite{Bhattacharyya:2018bbv}.
For the case of a free field theory one can follow the methods  developed in \cite{Jefferson:2017sdb} which allows to compute the complexity in the case where the reference state and the target states are Gaussian (as is the case for the vacuum of a free field theory).
A Gaussian state can be  characterized in normal coordinates by  a set of characteristic frequencies $\omega_k$; if the reference state is taken to have  all frequencies equal to a constant $\omega_0$, the complexity is given by
\begin{equation} \label{oscill-compl}
{\cal C} = \frac{1}{2} \, \sum_k \left|\ln \left({\omega_k \over \omega_0} \right) \right| \,.
\end{equation}
This formula is obtained as a geodesic distance, calculated in a certain metric defined on the space of unitary operators that are used to move within the
set of Gaussian states.
An essentially equivalent result was obtained in \cite{Chapman:2017rqy} via a different method, where the metric was computed from the Fubini-Study metric on the set of quantum states.\footnote{It has been shown in \cite{Guo:2018kzl} that the two methods will not be equivalent in general, and an explicit counterexample can be found using coherent states.}

In order to make a connection between eq.~\eqref{oscill-compl} and the holographic  model studied in the previous sections, we look at a free 1+1-dimensional CFT with a conformal defect, i.e., a defect which preserves at least one copy of the Virasoro algebra. A single defect on the real line can be mapped to a boundary using the ``folding trick'', and the problem of constructing conformally invariant boundary conditions has been considered by Cardy \cite{Cardy:1989ir} who derived a set of consistency conditions that boundary states have to satisfy. However these conditions cannot be solved in full generality.
In the simplest case of a single free boson, which we denote as $\phi_+$ and $\phi_-$ on the right and left sides of the defect respectively, it is possible to show \cite{Bachas:2001vj} that  the most general current-preserving conformal boundary condition relating the derivatives of the fields is
\begin{align}
&\left(\begin{matrix}
\del_x \phi_-  \\
\del_t \phi_-
\end{matrix}\right)=
M(\lambda)
\left(\begin{matrix}
\del_x \phi_+  \\
\del_t \phi_+
\end{matrix}\right), \qquad
M(\lambda)=\left(\begin{matrix}
\lambda & 0\\
0 & \lambda^{-1}
\end{matrix}\right), \quad \text{or}
\\
& \left(\begin{matrix}
\del_x \phi_-  \\
\del_t \phi_-
\end{matrix}\right)=
M'(\lambda)
\left(\begin{matrix}
\del_x \phi_+  \\
\del_t \phi_+
\end{matrix}\right),
\qquad
M'(\lambda)=\left(\begin{matrix}
0 & \lambda^{-1}\\
\lambda & 0
\end{matrix}\right) \,.
\end{align}
If the boson is compact, the first condition amounts to a change of the compactification radius, with $\lambda = R_+/R_-$.
The second type of defect is related to the first type by a T-duality on one side of the defect, i.e, $\partial_\mu \phi_+ = \epsilon_{\mu\nu} \partial^\nu \tilde \phi_+$.
These boundary conditions can be obtained by requiring that energy is conserved (i.e., the stress tensor component $T_{xt}$ is continuous)  at the location of the defect.

\begin{figure}
\centering
\includegraphics[scale=0.9]{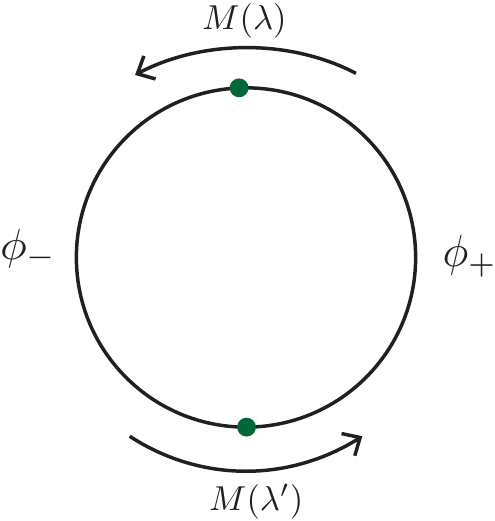} 
\caption{Illustration of two CFT domain walls at the opposite sides of a periodic domain.}
\label{DQFT}
\end{figure}

In order to mimic the setup of our holographic model we will consider a scalar field, living on a periodic boundary of length $2L_{\mathcal{B}}$, namely $x\in[-L_{\mathcal{B}},L_{\mathcal{B}}]$, with defects at the diametrically opposed points $x=0$ and $x=L_{\mathcal{B}}$, as indicated in figure \ref{DQFT}.

If the boson is compact, then its compactification radius must be unchanged after going once around the circle; this implies that the matrix associated to one defect must be the inverse of the other; if at $x=0$ we have a defect $M(\lambda)$, at the opposite side the defect has to be  $M(\lambda^{-1})$. This amounts to choosing $\lambda'=\lambda^{-1}$ in figure \ref{DQFT}.
It is a simple exercise to show that imposing these boundary conditions on the boson leads to a spectrum that is the same as in the theory without defects, namely
\begin{equation}
\omega_n = \frac{\pi n}{L_{\mathcal{B}}}\,.
\end{equation}
Applying  the formula \eqref{oscill-compl} then obviously leads to the result that the complexity does not depend on the presence of the defect. This is in agreement with the result for the CA conjecture that we obtained in \eqref{final-result-CA}, but not with the result for the CV conjecture \eqref{eq:CVresult}, which may be seen as an argument in favor of the CA conjecture. However, we should be cautious in drawing such a conclusion, as the model we consider is a very simple one and we do not know if the result is generic.
Moreover, the fact that the normal frequencies are the same does not imply that there is no effect of the defect. The zero modes play an important role  in determining the entanglement entropy, more precisely its finite part, which can be identified with the Affleck-Ludwig boundary entropy \cite{Sakai:2008tt,Gutperle:2017enx,Azeyanagi:2007qj}. This potentially hints that one has to incorporate the effect of zero modes, or winding modes, into the free field theory definitions of complexity \cite{Jefferson:2017sdb,Chapman:2017rqy}.

We can make a few further observations. First, notice that if the two defects are not placed at antipodal points, then the spectrum will change (the defects are not topological, so the theory depends on the distance between them). In the field theory we can put the defects wherever we want, but in the gravitational dual it seems difficult to find a corresponding solution where the brane would have to bend, so something would have to pull on it to stabilize the solution.
Second, if the boson is not compact, then there is no reason a priori why the two defects should be related to each other. We can allow for a more general pair of defects $M(\lambda), M(\lambda')$. The boundary conditions lead to a set of allowed momenta and corresponding frequencies
\begin{equation}\label{lambdalambdaprimefreq}
\omega_{n \pm}  = |k_{n\pm}|\,, \quad  k_{n\pm}=\frac{n \pi}{L_{\mathcal{B}}} \pm \frac{1}{L_{\mathcal{B}}}\tan^{-1}\left({\frac{\lambda\lambda'-1}{ \lambda +\lambda' }}\right) \equiv {\pi \over L_{\mathcal{B}}} (n \pm \Delta)\,.
\end{equation}
Alternatively, if one considers a pair of defects with transfer matrices $M(\lambda)$ and $M'(\lambda')$ the frequencies are again identical to those in the vacuum state, while for transfer matrices $M'(\lambda), M'(\lambda')$ we obtain the same result as in eq.~\eqref{lambdalambdaprimefreq}.

Even though we have no reason to think that this model has anything to do with our holographic model, we may hope that the corresponding complexity will have a sufficiently  generic form. Then using the prescription (\ref{oscill-compl}), and assuming for simplicity that we use a reference frequency larger than the cutoff, we find
\begin{equation} \label{two-defects-compl}
\begin{split}
 \mathcal{C} = &    \sum_{n=1}^{N}  \ln \frac{L_{\mathcal{B}}^2 \omega_0^2}{\pi^2 \left|n^2-\Delta^2\right|}  \sim  2 N \ln \left( {L_{\mathcal{B}} \omega_0 \over \pi N} \right)  +2N  -  \ln N -  \ln \left(\frac{2 \sin(\pi \Delta)}{ \Delta} \right)  \\
   &   =    2 \left(\frac{L_{\mathcal{B}} \Lambda}{ \pi }\right) \left[\ln \left( { \omega_0 \over \Lambda} \right)  +1\right]  -  \ln \left(\frac{L_{\mathcal{B}} \Lambda}{ \pi }\right)   -  \ln \left(\frac{2 \sin(\pi \Delta)}{ \Delta} \right) \,,
 \end{split}
\end{equation}
where we have used $N = L_{\mathcal{B}} \Lambda /\pi$ with $\Lambda$ the momentum UV cutoff. In the above expression we have not included the contribution of the mode $n=0$ since this mode would have to be IR regularized when considering the theory without the defect $\Delta=0$. Note that since we have assumed that $\omega_0>\Lambda$ the leading term in the complexity will be positive, as expected.

We can compare this result to the one of the holographic CV \eqref{eq:CVresult} and CA \eqref{final-result-CA} proposals. The field theory result has a  $\Lambda \ln \Lambda$ divergence, which is expected in field theory \cite{Chapman:2017rqy} but is absent both in CV and in CA when $\ell_{ct}$ is taken to be a constant. If we consider $\ell_{ct} \sim 1/\Lambda$ in the CA proposal, the leading divergence is reproduced, but not the subleading $\ln \Lambda$. The fact that the subleading divergences do not agree is not surprising given the simplicity of our model and was already observed in the complexity of the vacuum state in \cite{Jefferson:2017sdb,Chapman:2017rqy}. This choice of $\ell_{ct}$ would also lead to a $\ln^2 \Lambda$ divergence in the subregion complexity \eqref{toderive}, which despite being an unusual divergence to encounter in field theory quantities that need to be renormalized, does appear in quantum information measures, e.g., the entanglement and R\'enyi entropies for entangling surfaces which contains a conical singularity \cite{Bueno:2015lza}. Another option would be to choose the reference frequency $\omega_0 \sim \Lambda$. In this case the divergences are only $\Lambda$ and $\ln \Lambda$ and the structure is the same as for CV \eqref{eq:CVresult}, except for the fact that the coefficient of the log depends on the parameter of the defect in CV, whereas in the field theory the defect affects only the finite part. Comparing eq.~\eqref{two-defects-compl} to the results of the CA proposal \eqref{final-result-CA} we see that in both cases a defect dependent log contribution is absent.

It appears that the absence of a defect-dependent log is due to a cancelation that occurs between modes of momentum $k$ and $-k$; they are degenerate
in the free model, and the defect lifts the degeneracy symmetrically, i.e., $ \omega \to \omega \pm \delta\omega$. This suggests that the result can change if  parity invariance is broken, for instance in a chiral theory, or if the defect has degrees of freedom living on it; in this case there is a channel of  inelastic scattering of the modes, so that $k$ is not coupled only to $-k$.\footnote{Note, however, that the notion of degrees of freedom localized on the defect is not well-defined outside of the perturbative regime, see, e.g.,  \cite{Herzog:2017xha}.}
This idea can be checked explicitly in a solvable model  \cite{Callan:1994ub} of a free boson with a boundary interaction of the form
\begin{equation}\label{notinter}
L = \frac{1}{8\pi}  \int dx \left( \partial_\mu \phi \partial^\mu \phi \right)  - g  \cos\left({\phi(0) \over \sqrt{2}}\right) \,.
\end{equation}

The interaction term is of dimension one and is exactly marginal. By taking the boson at the self-dual radius, one can see that the interaction term can be reabsorbed into a redefinition of the SU(2) currents  $J^3 = {i \over \sqrt{2}} \partial \phi, J^\pm = e^{\pm i \sqrt{2} \phi}$:
\begin{equation}
J^1(x)  \to {\cal J}^1(x) = J^1(x) - \frac{g}{2} \delta(x) \,.
\end{equation}
The effect of this shift is to change the allowed $U(1)$ charges (i.e., momenta) of the modes. On a segment $[0,L_{\mathcal{B}}]$ with Dirichlet boundary conditions at $x=L_{\mathcal{B}}$, one finds that $k_n = \frac{\pi}{L_{\mathcal{B}}} (n + g/2)$; the complexity in this case gives a result similar to \eqref{two-defects-compl}, but with a term  $~ g \, \ln \Lambda$.
However, if the interaction term is added at both endpoints, the spectrum is different
\cite{Polchinski:1994my}: it forms continuous
bands centered around each integer, of width $1- 2 \alpha$ for $g \in \alpha+ {\mathbb Z}$. In this case the asymmetry disappears again.

\section{Discussion}\label{Discussion}

In this paper we have studied how the results of the holographic complexity proposals change when the boundary theory includes a conformal defect. We have focused on a simple gravity model which includes an AdS$_2$ brane embedded inside an AdS$_3$ geometry for which the full solution is known including backreaction \cite{Azeyanagi:2007qj}. The solution consists of two, slightly more than half, patches of empty AdS$_3$.

In section \ref{sec:CVCA} we evaluated  the complexity of the full boundary state according to the complexity=volume proposal and found that it has an additional logarithmic divergence, compared to the case of vacuum AdS$_3$, see eq.~\eqref{eq:CVresult}. We define the difference as the defect formation complexity
\begin{equation}\label{halfy}
\Delta C_{\text{defect}}^{\text{form}}\equiv \mC_V - \mC_{V,\text{vac}}
=\frac{8 c_T \, \sinh y^* }{3 } \ln\left(\frac{2  L_{\mathcal{B}}}{ \delta}\right) ,
\end{equation}
where we have used eq.~\eqref{eq:cutoff} to express $\hat \delta$ as a function of the UV-cutoff $\delta$ and the boundary radius $L_{\mathcal{B}}$. Since the coefficient of the logarithmic divergence does not depend on the regularization scheme, we expect that it is related to the physical data of our system. Indeed, we demonstrate below that in addition to the explicit dependence on the central charge, the coefficient of the logarithm in the above equation is related to the Affleck-Ludwig boundary entropy \cite{Affleck:1991tk}, which manifests itself as the finite part of the entanglement entropy \cite{Calabrese:2004eu} in the presence of a boundary.
In the case of the conformal defect studied in this paper, when the entanglement region is symmetric around the defect, it is possible to use a folding trick to relate the system with the defect to a finite system with boundaries and in this case the finite part of the entanglement entropy is also related to the boundary entropy. In the holographic setup when evaluating the entanglement \eqref{EEdefectsub} for an entangling region which is symmetric around the defect one obtains
\begin{equation}
S_{EE} = \frac{c_T}{3} \ln \left( \frac{2  L_{\mathcal{B}}}{ \delta} \sin\left(\frac{ \ell}{2 L_{\mathcal{B}}}\right)\right)+\ln g ; \qquad \ln g=\frac{c_T\, y^*}{3},
\end{equation}
where $\ell \equiv L_{\mathcal{B}} \Delta \theta$ is the length of the interval on the boundary and $\ln g$ is the boundary entropy. This result was already obtained in  \cite{Azeyanagi:2007qj} and eq.~\eqref{EEdefectsub} generalizes it to  the case of asymmetric regions around the defect; we find that in the latter case the finite part of the entanglement is no longer a constant but depends on the location of the endpoints. As far as we know, this result has not appeared in the literature before, and it would be interesting to have a field theory derivation of it.

Next, we evaluated the complexity according to the complexity=action proposal and found that the result was identical to the one obtained for empty AdS$_3$
\begin{equation}
\Delta C_{\text{defect}}^{\text{form}}=\mC_A - \mC_{A,\text{vac}}=0 .
\end{equation}
Previous studies of the two holographic proposals have found that the two results generally coincide up to an overall numerical factor.  This includes the late time growth rate of complexity in black hole backgrounds which is proportional to the mass of the black hole (for the CV proposal this is valid in the high temperature limit) \cite{Susskind:2014rva,Brown:2015bva,Brown:2015lvg},\footnote{See also \cite{Carmi:2017jqz} for the full time dependence.} characteristic delays in the complexity growth due to the introduction of shockwaves in the system
\cite{Stanford:2014jda,Zhao:2017iul,Chapman:2018dem,Chapman:2018lsv}, as well as the structure of divergences in holographic complexity (this is true when including a counter term in the action proposal), see \cite{Comments,Reynolds:2016rvl}. It is therefore interesting that for the case of the defect the results of the two holographic proposals dramatically differ. However, we have to be careful about the generality of this result. Holographic complexity is known to have special features for the case of $d=2$, e.g., the complexity of formation is not proportional to the entropy for this particular boundary dimension \cite{Formation}. It is therefore important to carry this analysis in higher dimensional cases before a definite conclusion can be drawn. One possibility would be to use the setup of AdS/BCFT to study the complexity in the presence of a boundary \cite{Takayanagi:2011zk,Fujita:2011fp} or in various holographic models proposed for systems with defects \cite{Bak:2003jk,Clark:2004sb,Karch:2000ct,DeWolfe:2001pq}. It would also be interesting to try and explore the effect of defects of codimension different than one on the complexity.  Another possible extension of our results would be to explore the effect of the defect at finite temperature when a black hole is present in the bulk.

It is important to point out that, at least naively, in order to study the complexity using the volume conjecture we have to include the backreaction of the brane. It would be interesting to see if this could somehow be avoided as was done for the entanglement entropy in \cite{Jensen:2013lxa} by using the Casini-Huerta-Myers trick \cite{Casini:2011kv}, or in an expansion for small tension of the brane, see \cite{Chang:2013mca}. Note that the analysis of the action cannot be performed in the probe approximation by simply considering the action of the brane itself since the gravitational action for the region surrounding the brane contributes at the same order in an expansion in the tension of the brane, as explicitly seen in our calculation.

In section \ref{sec:subregion} we evaluated the holographic complexity for subregions using the generalizations of the CA and CV proposals \cite{Alishahiha:2015rta,Carmi:2016wjl} motivated by the suggestion that the natural bulk region to encode the information about the reduced density matrix is the entanglement wedge \cite{Czech:2012bh,Headrick:2014cta}. Using the complexity=volume we found in eq.~\eqref{eq:subCVfinal} that  the leading divergence in the complexity was proportional to the size of the boundary region. This was already noted in \cite{Alishahiha:2015rta,Comments}. The defect introduced a subleading logarithmic divergence, which was exactly half of the one given in eq.~\eqref{halfy} for the full boundary state. The reason is that the subregion covers only one defect in the boundary theory.  It is also interesting to compare this result to the entanglement entropy in equation \eqref{EEdefectsub} where the contribution due to the defect was finite rather than logarithmic.

For the complexity=action for a symmetric region around the defect, we found again that the complexity with the defect was identical to the result for vacuum AdS$_3$, see eq.~\eqref{eq:subCAdresult}. We derived the result for empty AdS$_3$ in global coordinates in appendix \ref{app:sCAallpieces} and the final result can be found in eq.~\eqref{toderive} (previous results for subregions in vacuum AdS with a flat boundary can be found in \cite{Comments}). We observed that the leading divergence is proportional to the size of the interval and that a certain ambiguity was introduced by the parameter $\ell_{ct}$, with dimension of a length,
due to the gravitational counterterm needed to restore reparametrization invariance of the gravitational action. This counterterm was recently shown to be an essential ingredient in the CA proposal in order to reproduce certain desired properties of the complexity in the presence of shockwaves \cite{Chapman:2018dem,Chapman:2018lsv}. We also observed a subleading logarithmic divergence which depends on the same ambiguity due to the counterterm. If the characteristic length $\ell_{ct}$  is chosen to be of the order of the cutoff, this introduces a $\ln^2 \hat \delta$ divergence in the holographic complexity. It would be interesting to generalize this result to the case of a region which is not symmetric around the defect which cannot be related to a system with boundary using the folding trick.

In section \ref{sec:QFT} we studied the complexity of the ground state for two simple models of bosonic QFTs  including two defects at the two opposite sides of a periodic domain. We evaluated the complexity according to the methods introduced in \cite{Jefferson:2017sdb,Chapman:2017rqy} for Gaussian states in free quantum field theories, starting from an unentangled product state with characteristic frequency $\omega_0$.
The first model consists of a free boson with permeable domain wall defects. In this model, we found that the logarithmic contribution to the complexity does not depend on the permeability parameter $\lambda$ characterizing the defect, see eq.~\eqref{two-defects-compl}. This is similar to what happened in holography using the complexity=action proposal. Later, we considered an exactly solvable model with a boundary interaction given in eq.~\eqref{notinter}. In this model a logarithmic divergence which depends on the defect parameters appeared in the complexity, analogously to our result for the complexity=volume conjecture, although the logarithmic term is absent even in this case if the system has two boundaries.

No embedding of the precise holographic model studied in this paper into string theory is known. It would therefore be very interesting to reproduce our holographic calculation for a model that comes from a solution of string theory, for instance, one could consider the exact string background of AdS$_3$ with NS-NS fluxes, in which AdS$_2$ D-brane probes can be embedded, see \cite{Bachas:2000fr}. Presumably the fluxes would contribute to CA but not to CV, so there is a possibility that the discrepancy found in this paper would not be present in a full top-down model.
Another interesting possibility is to study the complexity in a smooth defect geometry, e.g., the Janus solution \cite{Bak:2007jm}, and check whether a defect-dependent logarithmic contribution is obtained using the CA/CV proposals.

It is possible to gain further intuition into the influence of the defect on the complexity by considering MERA circuits, for a review see \cite{2011arXiv1109.5334E}. MERA tensor networks constitute an efficient way of approximating the ground state of critical systems.
It has been suggested that they have a natural interpretation in holography where the MERA constitutes a lattice representation of a constant time slice in AdS and where the additional direction in the MERA circuit corresponds to the holographic RG scale \cite{Swingle:2009bg}. More precisely, the number of layers in the tensor network is proportional to $\log z$ where $z$ is the holographic FG coordinate. The lattice points in this description represent the MERA gates and so counting them (equivalently evaluating the volume of the time slice) would naturally result in a measure of the complexity of the state.

It was pointed out in \cite{Vidal:2007hda,Vidal:2008zz} that in order to find the ground state of a system whose Hamiltonian has been modified in a certain region due to an impurity or a defect it is sufficient to minimally update the tensor network, namely to replace the tensors in the causal cone of the defect, defined as the part of MERA which traces the evolution of the defect under coarse-graining transformations. Furthermore, if the defect is conformal, it is enough to replace the pair of tensors representing disentanglers and isometries with another (single) pair in the causal cone of the defect. For impurities spread over a small spatial region, the causal cone consists of approximately a fixed number of tensors at each layer.

We have seen in our CA calculations that the defect itself makes a large positive contribution logarithmic in the cutoff, see eq.~\eqref{CAdefects1234}, while the geometry around the defect introduces a negative contribution, which exactly cancels the one of the defect. This can be naturally interpreted in terms of the minimally updated MERA.
Introducing a cost for the tensors in the causal cone would give a log contribution, since it would be proportional to the length of the cone in the bulk; at the same time, we would have to subtract the contribution of the tensors that have been replaced. The exact cancellation that we observe seems to indicate that the CA proposal corresponds to microscopic rules where the defect gates and the ordinary gates are equally costly while in the CV, the defect gates are more costly.

An alternative interpretation was suggested in \cite{Czech:2016nxc} according to which the additional volume in the extension of the AdS space created by a thin defect could be interpreted as additional portions added to the tensor network, and this would resemble a discretized version of the time slice in our CV calculation. Another interesting possibility would be to incorporate a defect into the path integral complexity proposal based on Liouville action studied in \cite{Caputa:2017urj,Caputa:2017yrh,Caputa:2018kdj,Bhattacharyya:2018wym}.

We should point out that for entanglement entropy calculations in the free setup with a compact boson and two permeable domain walls, the zero modes play an important role; the finite boundary entanglement can be understood as arising essentially from the log of the volume of the zero modes \cite{Azeyanagi:2007qj}. This raises the question of whether the prescription for computing complexity using Gaussian states needs to be extended to account for a contribution of the zero modes. We leave this interesting issue for future study.

\section*{Acknowledgements}   
We would like to thank Alex Belin, Jan de Boer, Horacio Casini, Alejandra Castro, Bartek Czech,  Lorenzo Di Pietro, Ben Freivogel, Damian Galante, Andreas Karch, Marco Meineri, Juan Pedraza, Jan Troost and Erik Verlinde for many useful discussions. We would especially like to thank Davide Gaiotto, for suggesting this question to us, and Costas Bachas and Rob Myers, for many useful discussions and suggestions at various stages of this work and for sending us their comments on the manuscript. DG and GP would like to thank the hospitality of the Perimeter Institute for Theoretical Physics where part of this work was carried out. Research at Perimeter Institute is supported by the Government of Canada through the Department of Innovation, Science and Economic Development and by the Province of Ontario through the Ministry of Research, Innovation and Science. All the authors would like to thank the Galileo Galilei Institute for Theoretical Physics for hospitality during the workshop ``Entanglement in Quantum Systems'' where part of this work was carried out. SC would also like to thank the INFN for partial support during the workshop. SC acknowledges funding from the European Research Council (ERC) under starting grant No. 715656 (GenGeoHol) awarded to Diego M. Hofman.

\appendix

\section{Derivation of the Light Cone using Global Coordinates}\label{app:nullGeoApp}
In this appendix we derive the light cone surface of subsection \ref{sec:WDWpatch} using the $(\phi,\theta)$ coordinate system. Since the form of the null surface is identical in conformally equivalent spacetimes, we will study null geodesics in the metric
\begin{equation}\label{eqapp:metric.t.phi.theta.conf}
ds^2=-dt^2+d\phi^2+\sin^2 \phi \,\, d\theta^2
\end{equation}
which is a conformal rescaling of the metric \eqref{eq:metric.t.phi.theta}. The geodesic equations read,\footnote{Of course, since we are working with the equations of motion of the squared line element, under the conformal rescaling the parametrization of the null geodesics could change.}
\begin{equation}\label{eqapp:null.geo.eqs}
\ddot{t} = 0 ,\qquad
\ddot{\phi} = \frac{1}{2} \sin(2 \phi) \, \dot{\theta}^2 ,\qquad
\ddot{\theta} = -2 \cot\phi \, \dot{\theta} \, \dot{\phi} \, ,
\end{equation}
where the derivatives are taken with respect to some parameter $\sigma$ along the null geodesics and the requirement that the geodesics are null reads
\begin{equation}\label{eqapp:null.const}
-1 + {\dot{\phi}}^2 + \sin^2 \phi  \, {\dot\theta}^2 = 0.
\end{equation}
The first equation in \eqref{eqapp:null.geo.eqs} is consistent with having the geodesics parameterized by $t$, namely $\sigma=t$.
Integrating the last equation in \eqref{eqapp:null.geo.eqs} yields
\begin{equation}\label{eqapp:theta.dot.null.geo}
\dot{\theta} = \frac{a}{\sin^2 \phi},
\end{equation}
where $a$ is an integration constant. Next, we substitute the result from eq.~\eqref{eqapp:theta.dot.null.geo} into the null constraint \eqref{eqapp:null.const} which yields
\begin{equation}\label{eqapp:phi.dot.null.geo}
\dot{\phi} = \pm \sqrt{1-\frac{a^2}{\sin^2 \phi }}.
\end{equation}
Solutions to eqs.~\eqref{eqapp:theta.dot.null.geo}-\eqref{eqapp:phi.dot.null.geo} automatically satisfy the second equation in \eqref{eqapp:null.geo.eqs}. Before we proceed in finding the explicit solution for the null geodesics, let us pause and briefly comment on the properties of eqs.~\eqref{eqapp:theta.dot.null.geo}-\eqref{eqapp:phi.dot.null.geo}. If we start from the boundary and look at inwards and future oriented null rays we will choose the minus branch of eq.~\eqref{eqapp:phi.dot.null.geo} since  $\phi$ is decreasing as we move into the bulk. The constant of integration $a$ determines the angular orientation ($\theta$) of the null geodesic as it falls into the bulk according to eq.~\eqref{eqapp:theta.dot.null.geo}. For instance for $a=0$ we will have a geodesic which follows a line of constant $\theta$ and this is the geodesic which determines the boundary of the WDW patch in vacuum AdS$_3$ without the defect. For other values of $a$ there is a particular value of the radial coordinate $\phi=\sin^{-1}|a|$ for which $\dot{\phi}=0$ and the geodesic turns back toward the boundary.
Let us focus on the upper half of the WDW patch and study future oriented geodesics starting from the boundary.
Integrating the minus branch of the differential equation \eqref{eqapp:phi.dot.null.geo} we obtain the solution
\begin{equation}\label{eqapp:solt}
t = c_2 - \cos^{-1} \left( \frac{\cos \phi}{\sqrt{1-a^2}} \right) \qquad \text{where} \qquad \sin^{-1}|a|\leq\phi\leq\pi/2.
\end{equation}
The initial condition $\phi(t=0) = \pi/2$ fixes $c_2=\pi/2$. Finally we can solve the equation for $\dot\theta$, which gives
\begin{equation}\label{eqapp:c3c3}
\theta = c_3 +\tan^{-1} (a \tan t) \qquad \text{where} \qquad 0\leq t\leq\pi/2,
\end{equation}
and the initial condition $\theta(t=0) = 0$ for the null rays originating from $(t,\phi,\theta) = (0,\pi/2,0)$ fixes $c_3=0$.\footnote{For the null rays originating from $\theta(t=0) = \pi$ we have $c_3=\pi$ but since the picture is symmetric we will focus on the null rays originating from $\theta(t=0)=0$.}

We also present in the following an equation which describes the shape of the lightcone constructed from these null rays. We will focus on the part of the surface for negative angles $\theta<0$ which fixes the new component of the boundary of the WDW patch in the defect region, see figure \ref{fig:WDW}. First, from eq.~\eqref{eqapp:solt} with $c_2=\pi/2$ we extract $a$
\begin{equation}
a = - \sqrt{1-\frac{\cos^2 \phi}{\sin^2 t}} \qquad \text{for} \qquad \theta<0.
\end{equation}
Substituting this value into eq.~\eqref{eqapp:c3c3} together with $c_3=0$ for the null lightcone originating from the boundary at $\theta=0$ we obtain
\begin{align}
\tan\theta + \sqrt{\tan^2 t - \frac{\cos^2 \phi}{\cos^2 t}}=0 \qquad
\text{or} \qquad \cos \theta = {\cos t \over \sin \phi} \,.\label{lightconeaaa}
\end{align}
One can also relate this analysis to the $(y,r)$ coordinate one by verifying that lines of constant $a$ (or $t$) correspond to a constant value of the coordinate $y$ (or $r$) respectively, see eq.~\eqref{eq:coorconvert2}. More explicitly we have
\begin{equation}
a=\tanh y, \qquad
\tanh r =\cos t .
\end{equation}
A cross section of the light cone surface for different values of $t$ is depicted by the green slices in figure \ref{fig:ProfileWDWPatch} where we have used the following coordinates for the plot
\begin{equation}\label{eqapp:xycoords}
x=\frac{\phi}{\pi/2} \cos \theta, \qquad y=\frac{\phi}{\pi/2} \sin \theta.
\end{equation}

\section{Contributions to the Subregion CA Outside the Defect Region}\label{app:sCAallpieces}

In this appendix we evaluate the contributions to the subregion CA proposal outside the defect region for a subregion which is symmetric around the defect. This also gives the result for the subregion complexity from the CA proposal in empty AdS$_3$ in global coordinates for a subregion of the same size. The projections of the various regions used in this calculation onto the $t=0$ time slice are illustrated in figure \ref{fig:sCA}.
In what follows we will work in the $(t,\phi,\theta)$ coordinates of eq.~\eqref{eq:metric.t.phi.theta}. The relevant conversions can be found in eqs.~\eqref{eq:coordtrans}, \eqref{eq:coordtrans22} and \eqref{eq:coorconvert2}.

\paragraph{Preliminaries}
It will be useful to have expressions for the various surfaces in the relevant coordinates; these are: the boundary of the WDW patch ($S_3$ in figure \ref{fig:sCA})
\begin{equation}
\phi=\pi/2-t, 
\end{equation}
the boundary of the entanglement wedge, see eq.~\eqref{lightconeaaa} ($S_6 \cup S_7$ in figure \ref{fig:sCA})
\begin{equation}\label{app:BEW}
\sin \phi \cos \theta = \cos(\theta_R-t),
\end{equation}
the cutoff surface ($S_1 \cup S_5$ in figure \ref{fig:sCA})
\begin{equation}
\phi = \frac{\pi}{2}-\hat \delta, 
\end{equation}
the joint at the intersection of the WDW patch and the entanglement wedge ($J_4$ in figure  \ref{fig:sCA})
\begin{equation}
\cos \theta = \frac{\sin(\theta_R+\phi)}{\sin \phi} \qquad \text{and} \qquad
t=\pi/2-\phi=\tan^{-1}\left(\frac{\cos\theta -\cos\theta_R}{\sin\theta_R} \right),
\end{equation}
the point at the intersection of the joint $J_4$ and the $\theta=0$ surface
\begin{equation}
\phi=\frac{\pi}{2}-\frac{\theta_R}{2}, \qquad
t=\frac{\theta_R}{2} \qquad \text{and} \qquad \theta=0,
\end{equation}
while the point at the intersection of the joint $J_4$ and the cutoff surface is given by
\begin{equation}
t=\hat \delta, \qquad \phi = \pi/2-\hat\delta \qquad \text{and} \qquad \theta=\cos^{-1}\left(\frac{\cos (\theta_R-\hat \delta)}{\cos \hat \delta}\right)=\theta_R-\hat\delta+\mathcal{O}(\hat\delta^2).
\end{equation}
The RT surface $J_7$ is given by
\begin{equation}
t=0 \qquad \text{and} \qquad \sin \phi \cos \theta = \cos \theta_R,
\end{equation}
the point at the intersection of the RT surface and the surface $\theta=0$
\begin{equation}
\phi=\frac{\pi}{2}-\theta_R, \qquad
t=0\qquad \text{and} \qquad \theta=0,
\end{equation}
the joint  $J_1$ where the cutoff surface intersects the entanglement wedge
\begin{equation}\label{J1data}
\phi = \frac{\pi}{2}-\hat \delta, \qquad \cos \theta  = \frac{\cos(\theta_R-t)}{\cos \hat \delta},
\end{equation}
and finally, the point at the intersection of the RT surface and the cutoff
\begin{equation}\label{cutcut1234}
\theta=\cos^{-1}\left(\frac{\cos \theta_R}{\cos \hat \delta}\right)=\theta_R+\mathcal{O}(\hat\delta^2) \quad \text{and} \quad t=0.
\end{equation}

It will also be useful to have the surface data for the entanglement wedge in terms of the $(t,\phi,\theta)$ coordinates.
The null boundary of the entanglement wedge is given in eq.~\eqref{app:BEW} and can be parameterized similarly to eqs.~\eqref{eqapp:solt}-\eqref{eqapp:c3c3} with the substitution $t \rightarrow \theta_R-t$ in the relevant places
\begin{align}\label{eq:EWpara}
x^{\mu}=(t,\phi,\theta)=
\left(t,\cos^{-1} \left( \sqrt{1-a^2} \sin (\theta_R-t)\right), \tan^{-1} \left( a \tan (\theta_R-t)\right)\right),
\end{align}
where $a$ is constant along a given null geodesic parameterized by $t$. The surface information in this coordinate system with the parametrization $\lambda=t/\mathcal{N}_{\mt{EW}}$ is
\begin{align}
\begin{split}\label{appEWdatadata}
k^\mu_{\mt{EW}} &= \mathcal{N}_{\mt{EW}} \left(1, \frac{\sqrt{1-a^2}}{\sqrt{1+a^2 \tan^2 (\theta_R-t)}}, -  \frac{a\sec^2(\theta_R-t)}{1+a^2 \tan^2(\theta_R-t)} \right),\\
\kappa_{\mt{EW}}&=  2\mathcal{N}_{\mt{EW}} \cot(\theta_R-t),\qquad \gamma^{\mt{EW}}_{aa} = \frac{L^2}{(1-a^2)^2}.
\end{split}
\end{align}
We may also determine the values of $a$ and $t$ at the point where the entanglement wedge intersects RT surface and the cutoff surface
\begin{equation}\label{amax}
t=0, \qquad \text{and} \qquad a_{\text{max}}=\sqrt{1-\frac{\sin^2\hat \delta}{\sin^2\theta_R}}=1-  \frac{\hat{\delta}^2}{2\sin ^2\theta_R}  + \mathcal{O}(\hat \delta^4)
\end{equation}
as well as the point where it intersects the cutoff surface and the WDW patch
\begin{equation}\label{amin}
t=0, \qquad \text{and} \qquad a_{\text{min}}=\sqrt{1-\frac{\sin^2\hat \delta}{\sin^2(\theta_R-\hat \delta)}}=
1- \frac{\hat{\delta}^2}{2\sin ^2\theta_R}- \frac{\cot\theta_R }{\sin^2\theta_R}\hat{\delta}^3 + \mathcal{O}(\hat \delta^4).
\end{equation}

Below, we will only be able to extract the divergent pieces of the subregion CA proposal outside the defect region analytically and will leave some of the finite pieces as implicit integral expressions.

\paragraph{Bulk Contributions}
The bulk contribution $B_2$ under the WDW patch reads
\begin{equation}
\begin{split}
B_2 = &\,-\frac{L}{4 \pi \Gn} \,
 \int_{\frac{\pi}{2}-\frac{\theta_R}{2}}^{\frac{\pi}{2}-\hat \delta} d\phi\, \frac{\sin\phi}{\cos^3\phi}
 \int_0^{\cos^{-1}\left(\frac{\sin(\theta_R+\phi)}{\sin \phi}\right)} d\theta
 \int_0^{\frac{\pi}{2}-\phi} dt
\\
= &\,
  -{L \over 4\pi \Gn}\left( \frac{\theta_R}{\hat\delta} + \ln{\hat\delta}\right) + \text{finite}.
\end{split}
\end{equation}
Next, we evaluate the bulk contribution of the region under the entanglement wedge. We subdivide it into two parts,  $B_4$ and $B_5$, as indicated in figure \ref{fig:sCA}, along a line of constant $\phi = \frac{\pi}{2}-\frac{\theta_R}{2}$. The part $B_4$ is finite and reads
\begin{equation}
B_4 = -\frac{L}{4 \pi \Gn} \,
 \int_{\frac{\pi}{2}-\theta_R}^{\frac{\pi}{2}-\frac{\theta_R}{2}} d\phi\, \frac{\sin\phi}{\cos^3\phi}
 \int_0^{\cos^{-1}\left(\frac{\cos \theta_R}{\sin \phi}\right)} d\theta
 \int_0^{\theta_R-\cos^{-1} (\cos \theta \sin \phi)} dt
=\text{finite}.
\end{equation}
The part $B_5$ extends all the way to the cutoff and reads
\begin{equation}
\begin{split}
B_5 =&\, -\frac{L}{4 \pi \Gn} \,
 \int_{\frac{\pi}{2}-\frac{\theta_R}{2}}^{\frac{\pi}{2}-\hat \delta} d\phi\, \frac{\sin\phi}{\cos^3\phi}
 \int_{\cos^{-1}\left(\frac{\sin(\theta_R+\phi)}{\sin \phi}\right)} ^{\cos^{-1}\left(\frac{\cos \theta_R}{\sin \phi}\right)} d\theta
 \int_0^{\theta_R-\cos^{-1} (\cos \theta \sin \phi)} dt
\\
=&\,\frac{L}{8 \pi \Gn}\ln \hat \delta+\text{finite}.
\end{split}
\end{equation}
Therefore, the total bulk contribution reads
\begin{equation}\label{combo1}
I_{\text{bulk,out}}=-{L \over \pi \Gn}\left( \frac{\theta_R}{\hat\delta} + \frac{1}{2}\ln{\hat\delta}\right) + \text{finite},
\end{equation}
where we have included a factor of four to account for the two sides of the defect as well as the contributions above and below the $t=0$ time slice.

\paragraph{Surface Contributions}
We proceed by evaluating the various surface contributions. We start with the cutoff surface. In the region under the WDW patch ($S_1$), this is a simple modification of our previous calculation in eq.~\eqref{S1full}. All we have to do is modify the limits of integration as follows
\begin{equation}
S_1=
\frac{L}{8\pi\Gn} \int_0^{\hat\delta} dt \int_0^{\theta_R-\hat\delta+\mathcal{O}(\hat\delta^2)} d\theta \, \, \frac{\cos\hat{\delta}}{\sin^2\hat{\delta}} \left( \cos\hat{\delta} + \frac{1}{\cos\hat{\delta}}\right) =
\frac{L\,\theta_R}{4\pi \Gn\hat{\delta}} + \text{finite}.
\end{equation}
The part of the cutoff surface under the entanglement wedge is similarly given by
\begin{equation}
S_5=
\frac{L}{8\pi\Gn} \int_{\theta_R-\hat\delta+\mathcal{O}(\hat\delta^2)}^{\theta_R+\mathcal{O}(\hat\delta^2)} d\theta  \int_0^{\theta_R-\theta+\mathcal{O}(\hat\delta^2)} dt  \, \, \frac{\cos\hat{\delta}}{\sin^2\hat{\delta}} \left( \cos\hat{\delta} + \frac{1}{\cos\hat{\delta}}\right) = \text{finite}.
\end{equation}
Next, we evaluate the contributions of the various null surfaces.
We start with the null boundary of the WDW patch which requires a simple modification to the integration limits in eq.~\eqref{S3full12345}
\begin{align}
\begin{split}
S_3 &={L\over 8\pi\Gn} \int^{\theta_R -\hat{\delta}+\mathcal{O}(\hat \delta^2)}_0 d\theta \int_{\hat\delta}^{\tan^{-1}\left({\cos\theta -\cos\theta_R\over \sin\theta_R} \right)} \left(\frac{\ln \left(\frac{2 \ell_{ct} \mathcal{N}_3}{\sin(2t)}\right)+2\cos^2 t}{\sin^2 t}\right)\, dt \\
&
= {L\over 8\pi\Gn}\left({\theta_R \over \hat{\delta}}\left[ \ln \left( {\ell_{ct} \mathcal{N}_3 \over \hat{\delta}} \right) + 1 \right] + \ln \hat{\delta} \left[2+\ln(\ell_{ct}\mathcal{N}_3)\right]  - {1\over2}\ln^2 \hat{\delta}\right) + \text{finite}.
\end{split}
\end{align}
For the null boundary of the entanglement wedge, we will divide the integration region along a line of constant $a=a_\text{min}$, see eq.~\eqref{amin}, as indicated in figure \ref{fig:sCA}. This yields
\begin{equation}
S_6 = -\frac{L\mathcal{N}_{\mt{EW}}}{4\pi \Gn}\int_{a_{\text{min}}}^{a_{\text{max}}} da
\int_{0}^{\theta_R -\sin^{-1}{\left({\sin\hat\delta \over \sqrt{1-a^2}} \right)  }}   \frac{\cot(\theta_R-t)}{(1-a^2)} dt =\mathcal{O}(\hat \delta^2),
\end{equation}
and
\begin{equation}
S_7 = -\frac{L\mathcal{N}_{\mt{EW}}}{4\pi \Gn} \int_{0}^{a_{\text{min}}} da \int_{0}^{t_\text{max}}  \frac{\cot(\theta_R-t)}{(1-a^2)} dt = \text{finite} ,
\end{equation}
where $t_\text{max}$ solves the equation
\begin{equation}\label{tmax}
\sin t_\text{max} =\sqrt{1-a^2}\sin(\theta_R-t_\text{max}).
\end{equation}

\paragraph{Joint contributions}

The joint at the  intersection between the WDW patch and the cylindrical cutoff surface is similar to the expression \eqref{J1fullfull} and reads
\begin{align}
J_3
= -{L \over 8 \pi \Gn} \int_{0}^{\theta_R - \hat{\delta} +\mathcal{O}(\hat\delta^2) } \cot\hat{\delta}\ln \left(\frac{\mathcal{N}_3 L}{\sin\hat{\delta}}\right)d\theta
= -{L \over 8 \pi \Gn} \left( {\theta_R\over \hat\delta} -1\right)\ln\left({\mathcal{N}_3 L\over \hat\delta}\right).
\end{align}
For the joint at the intersection of the entanglement wedge and the cutoff surface we obtain
\begin{align}
J_1 &= -{1\over 8\pi \Gn}\int \sqrt{\gamma^{\mt{EW}}} \ln|k^\mu_{\mt{EW}} s_{\mu}^{(1)}| da\no\\
&= -{L\over 8\pi \Gn}\int_{a_{\text{min}}}^{a_{\text{max}}} \frac{1}{1-a^2}\ln \left( \frac{2 LN_{\mt{EW}} \sqrt{\cos^2 \hat \delta-a^2}}{\sin(2\hat \delta)}\right) da = \mathcal{O}(\hat\delta),
\end{align}
where we used eqs.~\eqref{eq:S1info1} and \eqref{appEWdatadata} for the relevant normal vectors.
Next, we evaluate the joint at the intersection of the entanglement wedge and the WDW patch. We use the normal vectors in eq.~\eqref{appEWdatadata} and \eqref{k3k3k3} and substitute the value of $t$ at the joint using eq.~\eqref{tmax}
\begin{align}
J_4
&=  \frac{L}{8\pi \Gn}\int_0^{a_{\text{min}}} \frac{1}{1-a^2} \left( \ln \frac{L^2 \mathcal{N}_3 \mathcal{N}_{\mt{EW}} \csc^2\theta_R}{2(1-a^2)} + \ln\frac{(2-a^2+2\sqrt{1-a^2}\cos\theta_R)^2}{1+\sqrt{1-a^2}\cos\theta_R}\right)da \no \\
&= \frac{L}{8\pi \Gn} \left(\ln ^2{\hat\delta} - \ln {\hat\delta} \ln \left({L^2 \mathcal{N}_3 \mathcal{N}_{\mt{EW}}\over  2}\right) \right) + \text{finite}.
\end{align}
Finally, the joint between the past and future boundaries of the entanglement wedge reads
\begin{align}
\begin{split}
J_7
&=  -\frac{L}{8\pi \Gn}\int_{0}^{a_{\text{max}}} \frac{da}{1-a^2} \ln \left(\frac{L^2 \mathcal{N}_{\mt{EW}}^2 \csc^2\theta_R}{1-a^2}\right) \\
&=- \frac{L}{8\pi \Gn}\left( \ln ^2{\hat\delta} - \ln {\hat\delta} \ln \left( L^2 \mathcal{N}_{\mt{EW}}^2\right)\right) + \text{finite}
.
\end{split}
\end{align}
Adding together the various joint and surface contributions yields
\begin{align}\label{combo2}
I_{\text{sj,out}}={L\over 2\pi\Gn}\left({\theta_R \over \hat{\delta}}\left[ \ln \left( {\ell_{ct}  \over L} \right) + 3 \right] + \ln \hat{\delta} \left[1+\ln\left(\frac{2\ell_{ct}}{L}\right)\right] \right) + \text{finite},
\end{align}
and of course we see that the various constant related to the choice of parametrization canceled out.

\paragraph{Total divergence}
Combining eqs.~\eqref{combo1} and \eqref{combo2}, we obtain the total divergence for the vacuum AdS$_3$ portion of the CA proposal for an entangling region which is symmetric around the defect
\begin{equation}
\hspace{-3pt}	\mathcal{C}_{A,\text{sub}}^\text{vac} = \frac{1}{\pi} \left(I_{\text{sj,out}}+I_{\text{bulk,out}}\right) =
{L\over 2\pi^2\Gn}\left({\theta_R \over \hat{\delta}}\left[ \ln \left( {\ell_{ct}  \over L} \right) + 1 \right] + \ln \hat{\delta} \,\ln\left(\frac{2\ell_{ct}}{L}\right) \right)
 + \text{finite}.
\end{equation}
This concludes the derivation of eq.~\eqref{toderive}.

\section{Subregion CV in the Poincar\'e patch}\label{app:poincare}
\begin{figure}
\centering
\includegraphics[scale=1]{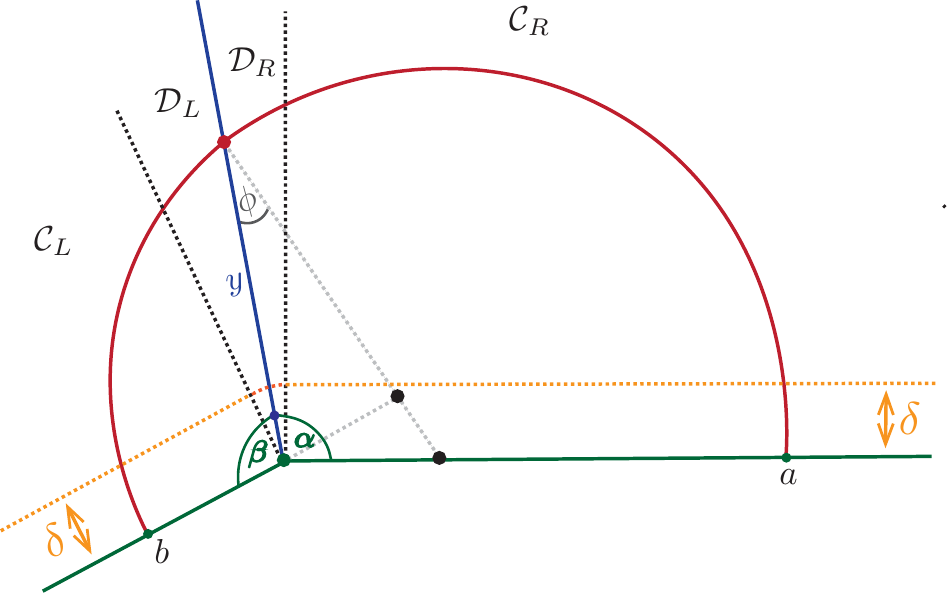}
\caption{Constant time slice of the two AdS$_3$ patches, glued together along the straight blue line representing the defect. $\alpha$ (right) and $\beta$ (left) are the opening angles of the patches on each side of the defect. The defect extends the space and its contribution is encoded in the regions $\mathcal{D}_L,\mathcal{D}_R$, while $\mathcal{C}_L,\mathcal{C}_R$ are the parts of the AdS$_3$ spaces outside the defect extension. The cutoff is represented by a dotted orange curve and is extended as a circular arc in the region of the defect. The red curve represents the RT surface corresponding to the region anchored at the AdS boundary at points $a$ and $b$. It consists of two circular arcs of different radii which are connected smoothly at the location of the defect. The centers of the relevant nested circles are depicted as black dots.} 
\label{fig:subCVPoincare}
\end{figure}

In the global coordinates, we are constrained to consider the case where the two patches glued together share the same cosmological constant. However, if we consider the Poincar\'e patch, we can glue together two AdS$_3$ patches with different AdS radii along the location of the defect.  The backreacted metric reads \cite{Bachas:2002nz,Czech:2016nxc,Erdmenger:2014xya}
\begin{align}
\hspace{-10pt} ds^2 =   {L^2 \over z^{\prime 2}}(-dt^2 + dz^{\prime 2} + dx^{\prime 2})\,\Theta( z^\prime + x^\prime\tan \beta)+{R^2 \over z^2}(-dt^2 + dz^2 + dx^2)\,\Theta(z - x\tan \alpha),
\end{align}
where $\Theta(z)$ is the Heaviside theta function and the two patches have different AdS radii $R$ and $L$. The matching condition is given by
\begin{equation}\label{eq:matchcondi}
\frac{R}{\sin\alpha} = \frac{L}{\sin\beta} = -\frac{\cot\alpha+\cot\beta}{8\pi \Gn \lambda},
\end{equation}
where $\lambda>0$ is the tension of the brane, see eq.~\eqref{eq:gravaction}. Stability of the gravitational solution requires that $\pi\geq \alpha,\beta\geq \pi/2$, see \cite{Czech:2016nxc}.
In the case with no defect ($\lambda =  0$) the matching condition \eqref{eq:matchcondi} implies that the two AdS radii are equal $L=R$ and that $\alpha = \beta = \pi/2$. However note that in general $L=R$ does not imply that there is no defect. We will think of the  two patches as being drawn alongside each other, see figure \ref{fig:subCVPoincare}. In this case the two coordinates systems $(t,x,z )$, and $(t,x^\prime,z^\prime )$ are related through a rotation in the $x-z$ plane as follows
\begin{align}
\begin{pmatrix}
z^\prime\\ x^\prime
\end{pmatrix}
=
\begin{pmatrix}
-\cos(\alpha+\beta) & \sin(\alpha + \beta)\\
-\sin(\alpha + \beta) & -\cos(\alpha+\beta)
\end{pmatrix}
\begin{pmatrix}
z\\x
\end{pmatrix}.
\end{align}
Figure \ref{fig:subCVPoincare} illustrates a constant time slice of our setup. The regions $\mathcal{D}_L$ and $\mathcal{D}_R$ are extensions of the AdS space due to the existence of the defect. We will consider the subregion complexity for a subregion which is anchored at the boundary at $x=a$ (right) and $x'=-b$ (left). Without loss of generality, we will everywhere assume that $a>b>0$.

It is natural to continue to choose the cutoff in the regions $\mathcal{C}_L$ and $\mathcal{C}_R$, to match the usual Fefferman-Graham expansion, with $z' = \delta$ and $z=\delta$, respectively.  In the defect region we suggest to extend the cutoff surface smoothly using a circular arc, similarly to what was done in subsection \ref{subsec:cutoff}.\footnote{One could in principle consider choosing two different cutoffs for the FG expansions on each side of the defect, connected by an interpolating curve in the defect region. For example one such choice was presented in \cite{Bachas:2002nz} where the author considers cutoffs of constant $z$ everywhere (including the defect region). The relation between the two cutoffs is determined by holding the scale factor of the metric fixed. The resulting cutoff is continuous but not smooth across the interface. In our case, these choices would not influence the logarithmic (defect-dependent) contribution to the complexity which is expected to be universal, see section \ref{Discussion}.} Focusing on the defect extensions $\mathcal{D}_R$ and $\mathcal{D}_L$, it will be useful to define a radial coordinates in the $x-z$ plane as
\begin{equation}
x=\rho \cos \theta, \qquad z=\rho \sin \theta.
\end{equation}
In terms of these coordinates the metric reads
\begin{equation}
ds^2 = \frac{R^2}{\rho^2 \sin ^2 \theta} \left(-dt^2 + d\rho^2 + \rho^2 d\theta^2 \right),
\end{equation}
and the cutoff extension is given by
\begin{equation}
\rho=\delta.
\end{equation}

The RT surfaces on each side of the defect are parts of circular arcs in the $x-z$ coordinates which are perpendicular to the boundary. They can be described by the following equations
\begin{equation}
(x-O_A)^2+z^2 = R_A^2, \qquad (x'-O_B)^2+z'^2=R_B^2,
\end{equation}
where $O_A$ and $O_B$ indicate the positions of the centers of the circular arcs and $R_A$ and $R_B$ indicate their Radii in the right/left patches respectively. The matching condition across the defect indicates that the circles are tangent at the location of the defect (see, e.g., \cite{Czech:2016nxc}). One is then led to the conclusion that the two centers as well as the location where the arcs intersect the defect lie along a single line. We define the angle between this line and the defect as $\phi$, see figure \ref{fig:subCVPoincare}.

The geometry is completely fixed by the angles $\alpha$ and $\beta$ as well as the sizes of the relevant boundary regions $a$ (right) and $b$ (left), see figure \ref{fig:subCVPoincare}. The angle $\phi$ is given by the solution to the equation
\begin{equation}
\frac{a}{b} =\left( \frac{\tan \frac{\alpha}{2} \tan \frac{\phi}{2}+1}{\tan \frac{\alpha}{2} \tan \frac{\phi}{2}-1} \right)
\left(\frac{\tan \frac{\beta}{2} \tan \frac{\phi}{2}+1}{\tan \frac{\beta}{2} \tan \frac{\phi}{2}-1}\right)
\end{equation}
which satisfies $\sin \alpha>\sin \phi$ and $\sin \beta> \sin \phi$, or explicitly
\begin{equation}\label{thetasol}
\tan\frac{\phi}{2} = \frac{(a+b)\sin \frac{\alpha+\beta}{2}-\sqrt{(a+b)^2\sin^2 \left(\frac{\alpha-\beta}{2}\right)+4 a b \sin\alpha \sin\beta}}{2 (a-b)\sin \frac{\alpha}{2} \sin \frac{\beta}{2}}.
\end{equation}
The various ``lengths''\footnote{Here we indicate the length without the overall conformal factor namely, $\sqrt{x^2+z^2}$ or $\sqrt{x'^2+z'^2}$.} indicated in figure \ref{fig:subCVPoincare} are given by
\begin{equation}
\begin{split}
R_A = a \frac{\sin \alpha}{\sin \alpha+\sin \phi}, \qquad R_B = b \frac{\sin \beta}{\sin \beta-\sin \phi},
\\
O_A = a \frac{\sin \phi}{\sin \alpha+\sin \phi}, \qquad O_B = b \frac{\sin \phi}{\sin \beta-\sin \phi},
\end{split}
\end{equation}
and the ``length'' along the defect up to the meeting point is given by
\begin{equation}
y=a\, \frac{\sin (\alpha+\phi)}{\sin \alpha+\sin \phi} = b\, \frac{\sin (\beta-\phi)}{\sin \beta-\sin \phi},
\end{equation}
or explicitly
\begin{equation}\label{soly}
y= \frac{(a-b)\sin \left({\beta-\alpha \over 2}\right) + \sqrt {(a+b)^2 \sin^2 \left({\beta-\alpha \over 2}\right)  +4ab \sin\alpha \sin\beta} }{2 \sin\left({\alpha+\beta \over2}\right)},
\end{equation}
cf.~eq.~(B.3) of \cite{Czech:2016nxc}.

We will focus on the subregion complexity, since the boundary is infinite and hence the complexity of the full space is (IR) divergent.
We have divided the integration region into four regions -- $\mathcal{C}_R$ and  $\mathcal{C}_L$ outside the defect region and  $\mathcal{D}_R$ and  $\mathcal{D}_L$ inside the defect region. The various volumes read
\begin{equation}
\begin{split}
\mathcal{C}_R =&\,  R^2 \int_\delta^{\sqrt{R_A^2-O_A^2}} \frac{dz}{z^2} \int_0^{O_A + \sqrt{R_A^2 -z^2}} dx
+R^2 \int_{\sqrt{R_A^2-O_A^2}}^{R_A} \frac{dz}{z^2} \int_{O_A-\sqrt{R_A^2-z^2}}^{O_A+\sqrt{R_A^2-z^2}}dx
\\
=&\, R^2 \left(\frac{a}{\delta}  -\frac{\pi}{2}-\sin^{-1} \left(\frac{\sin \phi}{\sin \beta}\right)\right),
\end{split}
\end{equation}
and
\begin{equation}
\begin{split}
\mathcal{C}_L = L^2 \int_\delta^{\sqrt{R_B^2-O_B^2}} \frac{dz'}{z'{}^2} \int^0_{O_B - \sqrt{R_B^2 -z'{}^2}} dx'
= L^2 \left(\frac{b}{\delta} -\frac{\pi}{2}+\sin^{-1} \left(\frac{\sin \phi}{\sin \beta}\right)\right),
\end{split}
\end{equation}
outside the defect region and
\begin{equation}
\begin{split}
\mathcal{D}_R =  R^2 \int_{\pi/2}^\alpha \frac{d\theta}{\sin^2 \theta} \int_\delta^{O_A \cos\theta +\sqrt{R_A^2-O_A^2 \sin^2\theta}} \frac{d\rho}{\rho}= R^2 \left(\cot \alpha \ln \left(\frac{\delta}{y}\right)-\phi+\sin^{-1} \left(\frac{\sin\phi}{\sin\alpha} \right) \right),
\end{split}
\end{equation}
and
\begin{equation}
\begin{split}
\mathcal{D}_L =  L^2 \int^{\pi/2}_{\pi-\beta} \frac{d\theta}{\sin^2 \theta} \int_\delta^{O_B \cos\theta +\sqrt{R_B^2-O_B^2 \sin^2\theta}} \frac{d\rho}{\rho}
=L^2 \left(\cot \beta \ln \left(\frac{\delta}{y}\right)+\phi-\sin^{-1} \left(\frac{\sin\phi}{\sin\beta} \right) \right),
\end{split}
\end{equation}
inside the defect region.
Of course we notice that the left and right patch results are related by the exchange $\phi \rightarrow -\phi$, $\alpha \rightarrow \beta$, $a \rightarrow b$ and $R\rightarrow L$.

Summing all the contributions together, multiplying by the relevant factors of proportionality $1/(\Gn R)$ and $1/(\Gn L)$ for the right and left patches respectively, see eq.~\eqref{eq:CVformula}, and expressing the result in terms of the central charges, $c_R \equiv \frac{3R}{2 \Gn}$ and $c_L \equiv \frac{3L}{2 \Gn}$, yields
\begin{equation}\label{resultofapc}
\begin{split}
\mathcal{C}_V^\text{poincare} = &\,\frac{2(c_R a+c_L b)}{3\delta}+\frac{2}{3}
\left(c_R \cot \alpha + c_L \cot \beta \right)\ln \left(\frac{\delta}{y}\right)
\\
&\,
-\frac{2 c_R}{3}  \left(
\frac{\pi}{2}+\phi \right)
 -\frac{2c_L}{3}   \left(
 \frac{\pi}{2}-\phi \right).
\end{split}
\end{equation}
where $\phi$ is defined in eq.~\eqref{thetasol}, $y$ is defined in eq.~\eqref{soly} and $\alpha$ and $\beta$ are determined from the matching conditions \eqref{eq:matchcondi} in terms of the two cosmological constants and the data of the defect.

Finally, let us study the limit of equal cosmological constant which yields $\alpha=\beta$, $c_R=c_L\equiv c_T$ and as a consequence $y=\sqrt{ab}$. This yields the following complexity
\begin{equation}
\mathcal{C}_V^\text{poincare} = \frac{2c_T}{3} \left( \frac{a+ b}{\delta}-
2  \sinh y^* \ln \left(\frac{\delta}{\sqrt{ab}}\right) -\pi \right)
\end{equation}
where we have used eq.~\eqref{eq:matchcondi} and eq.~\eqref{eq:defparam} to relate $\cot\alpha= -\sinh y^*$.
This can be seen as the large boundary size $a,b\ll L_\mathcal{B}$ limit of the global coordinate result of eq.~\eqref{eq:subCVfinal} where we have used $\theta_R = a/L_\mathcal{B}$ and  $\theta_L = -b/L_\mathcal{B}$ in order to perform the expansion in eqs.~\eqref{thetaLthetaR} and \eqref{eq:subCVfinal}.

A similar analysis to the one we have performed in subsection \ref{subcasecsec} for the subregion CA conjecture can be performed here as well. We leave this extension for future work.

\bibliographystyle{JHEP}
\bibliography{bibCD}{}

\end{document}